\documentclass[longauth]{aa}
\usepackage{color}
\usepackage{natbib}
\bibpunct{(}{)}{;}{a}{}{,}

\newcommand{\subsc}[1]{\ensuremath{_{\textrm{\scriptsize #1}}}}

\newcommand\capion[2]{\mbox{#1\hspace{0.2em}{\rmfamily\@Roman{#2}}}}

\newcommand{\Vmax}{\ensuremath{V_{\mathrm{max}}}}
\newcommand{\Diso}{\ensuremath{D_{\mathrm{iso}}}}
\newcommand{\isoA}{\ensuremath{\mathit{isoA}_{\mathrm{r}}}}
\newcommand{\isoB}{\ensuremath{\mathit{isoB}_{\mathrm{r}}}}
\newcommand{\Msun}{\ensuremath{\mathrm{M}_{\odot}}}

\begin{document}

\title{CALIFA: a diameter-selected sample for an integral field spectroscopy galaxy survey\thanks{Based on observations collected at the Centro Astron\'omico Hispano Alem\'an (CAHA) at Calar Alto, operated jointly by the Max Planck Institute for Astronomy and the Instituto de Astrof\'isica de Andaluc\'ia (CSIC).}}
\subtitle{}

\author{C.J.~Walcher\inst{\ref{i:AIP}}; L.~Wisotzki\inst{\ref{i:AIP}}; S.~Bekerait{\'e}\inst{\ref{i:AIP}}; B.~Husemann\inst{\ref{i:AIP},\ref{i:ESO}}; J.~Iglesias-P\'aramo\inst{\ref{i:IAA},\ref{i:CAHA}}; N.~Backsmann\inst{\ref{i:AIP}}; J.~Barrera Ballesteros\inst{\ref{i:IAC},\ref{i:UniLaguna}}; C.~Catal\'an-Torrecilla\inst{\ref{i:UnMad}}; C.~Cortijo\inst{\ref{i:IAA}}; A.~del Olmo\inst{\ref{i:IAA}}; B.~Garcia Lorenzo\inst{\ref{i:IAC},\ref{i:UniLaguna}}; J.~Falc{\'o}n-Barroso\inst{\ref{i:IAC},\ref{i:UniLaguna}}; L.~Jilkova\inst{\ref{i:UniPrague}}; V.~Kalinova\inst{\ref{i:MPIA}}; D.~Mast\inst{\ref{i:CAHA},\ref{i:rio}}; R.A.~Marino\inst{\ref{i:UnMad}}; J.~M\'endez-Abreu\inst{\ref{i:IAC},\ref{i:UniLaguna},\ref{i:andrews}}; A.~Pasquali\inst{\ref{i:zah}}; S.F.~S\'anchez\inst{\ref{i:IAA},\ref{i:CAHA},\ref{unam}}; S.~Trager\inst{\ref{i:Kapteyn}}; S.~Zibetti\inst{\ref{i:florence}}; J.A.L.~Aguerri\inst{\ref{i:IAC},\ref{i:UniLaguna}}; Alves, J.\inst{\ref{i:vienna}}; J.~Bland-Hawthorn\inst{\ref{i:SydUni}}; A.~Boselli\inst{\ref{i:LAM}}; A.~Castillo Morales\inst{\ref{i:UnMad}}; R.~Cid Fernandes\inst{\ref{i:cid}}; H.~Flores\inst{\ref{i:ObsParis}}; L.~Galbany\inst{\ref{i:UnChil1},\ref{i:UnChil2}}; A.~Gallazzi\inst{\ref{i:florence},\ref{i:dark}}; R.~Garc\'{i}a-Benito\inst{\ref{i:IAA}}; A.~Gil de Paz\inst{\ref{i:UnMad}}; R.M.~Gonz{\'a}lez-Delgado\inst{\ref{i:IAA}}; K.~Jahnke\inst{\ref{i:MPIA}}; B.~Jungwiert\inst{\ref{i:Czech}}; C.~Kehrig\inst{\ref{i:IAA}}; M.~Lyubenova\inst{\ref{i:MPIA},\ref{i:Kapteyn}}; I.~M{\'a}rquez Perez\inst{\ref{i:IAA}}; J.~Masegosa\inst{\ref{i:IAA}}; A.~Monreal Ibero\inst{\ref{i:AIP},\ref{i:ObsParis}}; E.~P{\'e}rez\inst{\ref{i:IAA}}; A.~Quirrenbach\inst{\ref{i:zah}}; F.F.~Rosales-Ortega\inst{\ref{i:mexico}};  M.M.~Roth\inst{\ref{i:AIP}}; P.~Sanchez-Blazquez\inst{\ref{i:uam}}; K.~Spekkens\inst{\ref{i:can}}; E.~Tundo\inst{\ref{i:florence}}; G.~van de Ven\inst{\ref{i:MPIA}}; M.A.W.~Verheijen\inst{\ref{i:Kapteyn}}; J.V.~Vilchez\inst{\ref{i:IAA}}; B.~Ziegler\inst{\ref{i:vienna}}}

\institute{
Leibniz-Institut f\"ur Astrophysik Potsdam (AIP), An der Sternwarte 16, D-14482 Potsdam, Germany\label{i:AIP} 
\and
European Southern Observatory, Karl-Schwarzschild-Str. 2, 85748 Garching b. M\"unchen, Germany \label{i:ESO}
\and
Instituto de Astrof\'{i}sica de Andaluc\'{i}a (CSIC), Glorieta de la Astronom\'{\i}a, s/n, 18008 Granada, Spain \label{i:IAA}
\and
Centro Astron\'{o}mico Hispano Alem\'{a}n, Calar Alto, (CSIC-MPG), C/Jes\'{u}s Durb\'{a}n Rem\'{o}n 2-2, E-04004 Almer\'{\i}a, Spain\label{i:CAHA}
\and
Instituto de Astrof\'{i}sica de Canarias, C/V\'{\i}a L\'actea S/N, 38200-La Laguna, Tenerife, Spain\label{i:IAC}
\and
Dept. Astrof\'{i}sica, Universidad de La Laguna, C/ Astrof\'{i}sico Francisco S\'anchez, E-38205-La Laguna, Tenerife, Spain\label{i:UniLaguna}
\and
Departamento de Astrof\'{i}sica y CC. de la Atm\'{o}sfera, Universidad Complutense de Madrid, Madrid 28040, Spain\label{i:UnMad}
\and
Department of Theoretical Physics and Astrophysics, Faculty of Science, Masaryk University, Kotl{\'a}rsk{\'a} 2, CZ-611 37 Brno, Czech Republic \label{i:UniPrague}
\and
Max Planck Institute for Astronomy, K\"onigstuhl 17, D-69117 Heidelberg, Germany \label{i:MPIA}
\and
Instituto de Cosmologia, Relatividade e Astrof\'{i}sica -- ICRA, Centro Brasileiro de Pesquisas F\'{i}sicas, Rua Dr.~Xavier Sigaud 150, CEP 22290-180, Rio de Janeiro, RJ, Brazil \label{i:rio}
\and
School of Physics and Astronomy, University of St Andrews, North Haugh, St Andrews, KY16 9SS, UK \label{i:andrews}
\and 
Zentrum f\"ur Astronomie der Universit\"at Heidelberg, Astronomisches Recheninstitut, M\"onchhofstr. 12 - 14, 69120 Heidelberg, Germany, \label{i:zah}
\and 
Instituto de Astronom\'\i a, Universidad Nacional Auton\'oma de Mexico, A.P. 70-264, 04510, M\'exico,D.F.  \label{unam}
\and
Kapteyn Astronomical Institute, Rijksuniversiteit Groningen, Postbus 800, NL-9700 AV Groningen, The Netherlands \label{i:Kapteyn}
\and
INAF-Osservatorio Astrofisico di Arcetri, Largo Enrico Fermi 5, I-50125 Firenze, Italy \label{i:florence}
\and
University of Vienna, T\"urkenschanzstr. 17, 1180 Vienna, Austria \label{i:vienna}
\and
Institute of Astronomy, School of Physics, University of Sydney, NSW 2006, Australia \label{i:SydUni}
\and
Laboratoire d'Astrophysique de Marseille - LAM, Universit\'e d'Aix-Marseille \& CNRS, UMR7326, 38 rue F. Joliot-Curie, F-13388 Marseille Cedex 13, France \label{i:LAM}
\and
Departamento de F\'{i}sica, Universidade Federal de Santa Catarina, PO Box 476, 88040-900, Florian\'{o}polis, SC, Brazil\label{i:cid}
\and
Laboratoire GEPI, Observatoire de Paris, CNRS-UMR8111, Univ Paris Diderot, 5 place Jules Janssen, 92195 Meudon, France \label{i:ObsParis}
\and
Millennium Institute of Astrophysics, Universidad de Chile, Casilla 36-D, Santiago, Chile \label{i:UnChil1}
\and
Departamento de Astronom\'ia, Universidad de Chile, Casilla 36-D, Santiago, Chile \label{i:UnChil2}
\and
Dark Cosmology Center, Niels Bohr Institute, University of Copenhagen, Juliane Maries Vej 30, 2100 Copenhagen, Denmark \label{i:dark}
\and
Astronomical Institute, Academy of Sciences of the Czech Republic, Bo\v{c}n\'{i} II 1401/1a, CZ-141\,00 Prague, Czech Republic \label{i:Czech}
\and
Instituto Nacional de Astrof{\'i}sica, {\'O}ptica y Electr{\'o}nica, Luis E. Erro 1, 72840 Tonantzintla, Puebla, Mexico \label{i:mexico}
\and
Departamento de Fisica Teorica, Universidad Autonoma de Madrid, 28049, Spain \label{i:uam}
\and
Department of Physics, Royal Military College of Canada, P.O. Box 17000, Station Forces, Kingston, Ontario, Canada K7K 7B4 \label{i:can}
}

\authorrunning{CALIFA collaboration}
\titlerunning{The CALIFA sample}


\date{Received date / Accepted date }

\abstract{
We describe and discuss the selection procedure and statistical properties of the galaxy sample used by the Calar Alto Legacy Integral Field Area Survey (CALIFA), 
a public legacy survey of 600 galaxies using integral field spectroscopy. 
The CALIFA `mother sample' was selected from the 
Sloan Digital Sky Survey (SDSS) DR7 photometric catalogue to include all galaxies with an $r$-band isophotal major axis between $45$\arcsec~and 
$79.2$\arcsec~and with a redshift $0.005 < z < 0.03$. The mother sample contains 939 objects, 600 of which will be observed in the course of the CALIFA 
survey. The selection of targets for observations is based solely on visibility and thus keeps the statistical properties of the mother sample. By comparison 
with a large set of SDSS galaxies, we find that the CALIFA sample is representative of galaxies over a luminosity range of $-19 > M_{\mathrm{r}} > -23.1$ 
and over a stellar mass range between $10^{9.7}$ and $10^{11.4} \Msun$. In particular, within these ranges, the diameter selection does not lead to any 
significant bias against -- or in favour of -- intrinsically large or small galaxies. Only below luminosities of $M_{\mathrm{r}} = -19$ (or stellar masses 
$<10^{9.7} \Msun$) is there a prevalence of galaxies with larger isophotal sizes, especially of nearly edge-on late-type galaxies, but such galaxies form 
$<10$\% of the full sample. We estimate volume-corrected distribution functions in luminosities and sizes and show that these are statistically fully 
compatible with estimates from the full SDSS when accounting for large-scale structure. For full characterization of the sample, we also present a 
number of value-added quantities determined for the galaxies in the CALIFA sample. These include consistent multi-band photometry based on 
growth curve analyses; stellar masses; distances and quantities derived from these; morphological classifications; and an overview of available 
multi-wavelength photometric measurements. We also explore different ways of characterizing the environments of CALIFA galaxies, finding that 
the sample covers environmental conditions from the field to genuine clusters. We finally consider the expected incidence of active galactic nuclei 
among CALIFA galaxies given the existing pre-CALIFA data, finding that the final observed CALIFA sample will contain approximately 30 Sey2 galaxies.  }

\maketitle

\section{Introduction}
\label{s:intro}

Spectroscopic surveys of galaxies are designed to helping understanding galaxy evolution by characterization of the properties of 
their targets. The two main physical properties of galaxies that are thought to drive galaxy evolution are galaxy mass and environment. 
All other processes that are very important for galaxies, such as active galactic nuclei (AGN), merging, gas accretion, and secular evolution,
should ultimately be consequences of these two characteristics, albeit with significant scatter. The dynamical time scales 
of large structures in the Universe are longer than a Hubble time and much longer than internal processes in galaxies, and 
therefore environmental effects do not have the time to average out. Surveys of large samples of galaxies are therefore 
needed to provide enough statistics in the presence of this scatter. 

In this paper we describe and discuss the target selection procedure for the Calar Alto Legacy Integral Field Area (CALIFA) 
Survey \citep{sanchez12a} and the resulting properties of the sample. CALIFA uses Integral Field Spectroscopy (IFS) to derive 
the spatial distributions of galaxy properties in two dimensions. The survey focusses on typical galaxies in the local Universe 
over a broad range of luminosities and types (yet avoiding dwarfs). For a more extensive description of the science case, we 
refer to \citet{sanchez12a}. 

Surveys are based on samples, which are constructed to represent populations. The ideal sample is volume-complete, 
i.e.\ it contains all galaxies within a given survey volume. In practice this is impossible to achieve (we still do not even know 
all galaxies in the Local Group) and can only, if at all, be approached by imposing substantial limits on the range of galaxy 
properties. For example, the ATLAS$^{3D}$ survey \citep{cappellari11} has been restricted to morphologically pre-classified early-type 
galaxies with $M_K < -18.5$ and thus managed to target an approximately volume-limited sample at distances $< 20$~Mpc. 
This would not have been possible for a more general survey, that includes later morphological types for which redshift-independent 
distance estimates are less complete. Any more general galaxy survey therefore needs to make selections based on some simple and 
accessible observational quantity, such as flux within a given filter band or a sufficiently precise definition of apparent size. 
While size selection of galaxies was very common in the days of visual scans of photographic atlases 
\citep{nilson73, davies90, de-vaucouleurs91}, the advent of digital imaging has shifted the focus towards favouring flux-limited 
surveys of galaxies \citep[e.g.][]{eisenstein01, strauss02, davis03, le-fevre05}. It is nevertheless useful to keep in mind that 
both selection methods (fluxes and sizes) are very similar in many aspects and that, in particular, the statistical methods of 
inferring population properties from observed samples are the same \citep[see][]{de-jong94}.

In this context it is useful to remind the reader that the consequences of `selection effects' may be entirely benign. Selection 
effect means that the statistical properties of a sample differ from those of the underlying population. However, selection effects 
can be corrected for in many cases, or taken into account by explicitly limiting the 
range where the sample is supposed to be representative. A bias arises only if the sample is devoid of certain types of 
objects that should be present, but are not in the sample, or if objects are underrepresented so an appropriate correction is not 
possible. The purpose of this paper is to understand the selection effects on the CALIFA mother sample in order to avoid 
biases. 

The instrument used for CALIFA is the Potsdam Multi-Aperture Spectrophotometer  \citep[PMAS,][]{roth05} 
mounted on the 3.5m telescope of the Calar Alto observatory, and employing the PPak wide-field integral field unit \citep[IFU,][]{kelz06} 
to sample a field of view (FoV) of $\sim 1$ arcmin$^2$. The PPak IFU was designed and custom-built for the DiskMass Survey,
which studies a size-selected sample of nearly face-on spiral galaxies based on isophotal diameters and signal-to-noise considerations
\citep{verheijen04, bershady10}. One of the major design drivers for the CALIFA sample selection was 
to take advantage of PPaK's large FoV and cover a large sample of galaxies of all types over their full optical extents.

However, observing a large sample of low-redshift galaxies with integral field spectroscopy in a homogeneous way is a 
challenge, because of the huge variations in the apparent sizes of galaxies. Any galaxy sample primarily defined 
by a selection cut on either apparent fluxes or intrinsic luminosities (or stellar masses) will invariably lead to a predominance 
of galaxies with small apparent sizes which significantly underfill the PPak IFU. For CALIFA we have chosen to follow a 
conceptually very simple approach, namely to directly select on angular isophotal sizes matched to the PMAS/PPak instrumental 
FoV. We decided to use isophotal sizes rather than Petrosian radii or some other size measure related to enclosed flux, 
because each isophote can be directly translated into an (approximately) constant minimal signal-to-noise (S/N) in the 
spectral continuum, as demonstrated by \citet[][specifically Sect.~6.5]{sanchez12a}.

CALIFA is conceived as a public legacy survey. The first set of data for 100 galaxies has already been released  \citep[DR1, ][]{husemann13}, 
and further data releases will follow. The present paper serves two purposes, both directed at present and future users of the CALIFA 
database. Firstly, we want to present the full information available about the CALIFA sample in a single place. 
And secondly, we wish to provide the users with an understanding of the usefulness and limitations of the sample to 
represent the galaxy population in the local Universe. Throughout this paper we use a cosmology defined by $H_0$ = 70 km/s/Mpc 
and $\Omega_{\Lambda}$ = 0.7 and a flat Universe.

\begin{figure*}[tbp]
\begin{center}
\includegraphics[width=0.49\hsize]{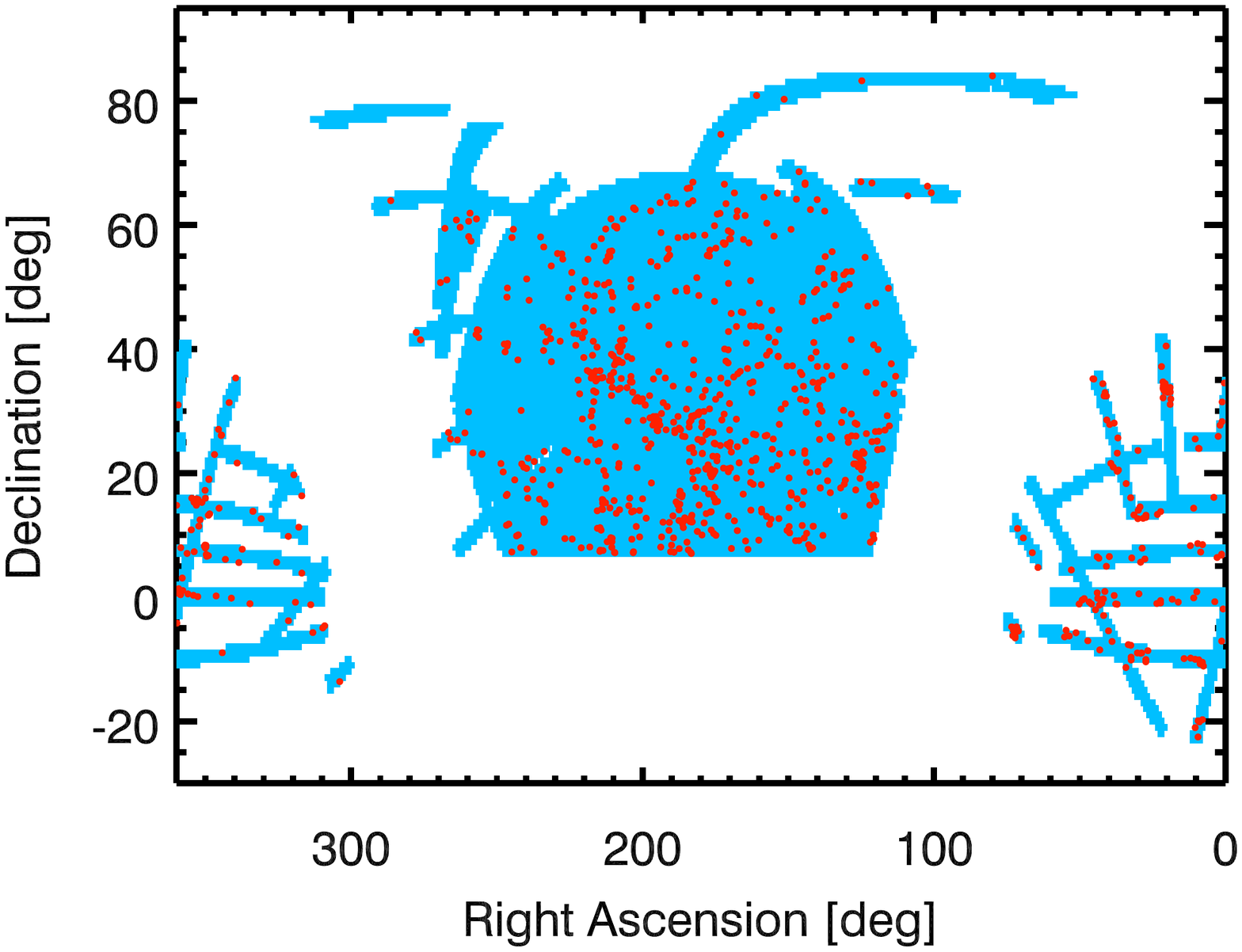}
\includegraphics[width=0.5\hsize]{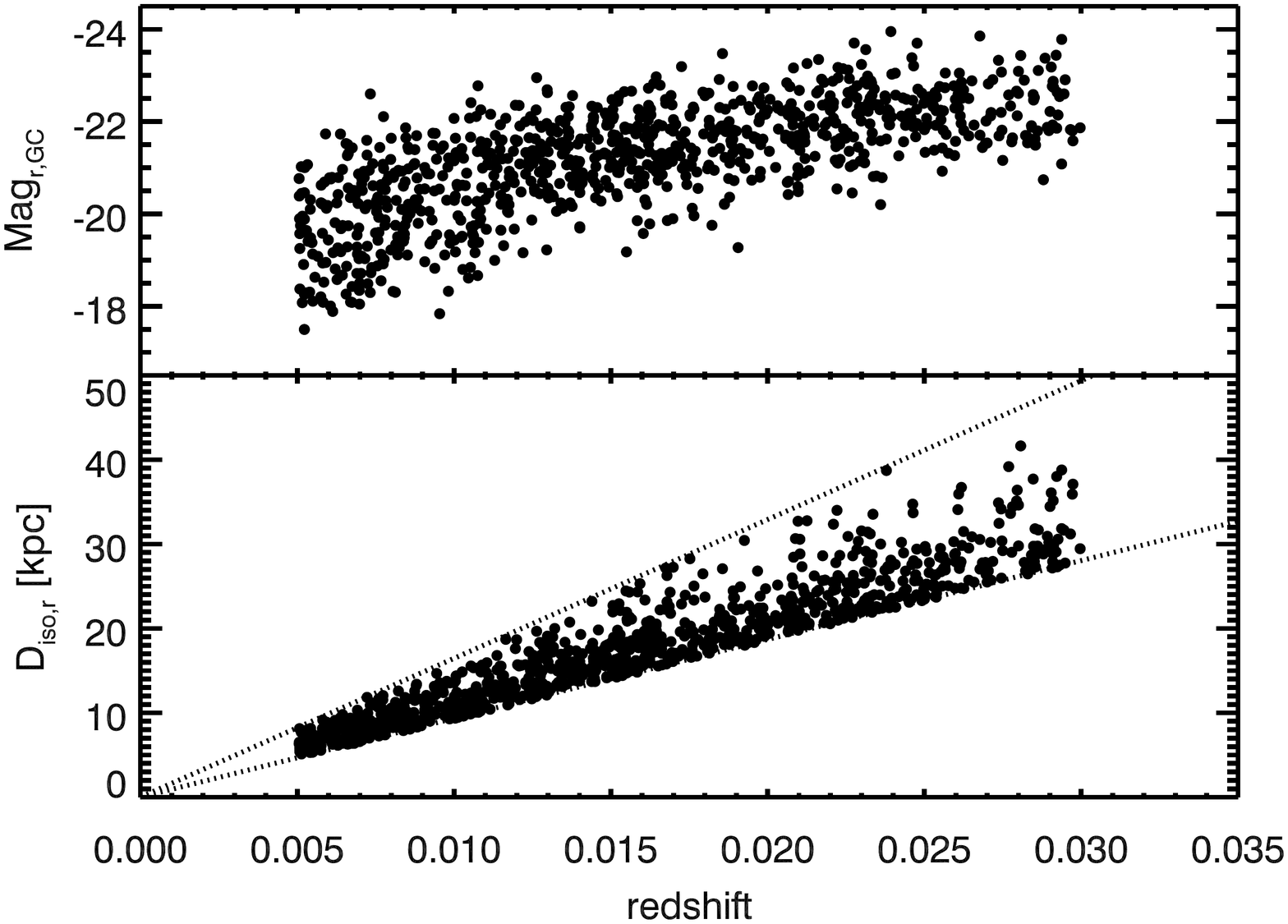}
\end{center}
\caption[geometry]{\emph{Left panel:} The footprint on the sky of our search in the DR7 CAS (light blue) and the distribution of the 
939 galaxies constituting the CALIFA mother sample (red circles). \emph{Right panels:} Redshifts vs.\ absolute magnitudes 
$M_{r,\mathrm{GC}}$ (top) and $r$-band linear isophotal sizes (bottom) for the galaxies in the sample.  The dotted lines 
in the lower panel are the selection limits.}
\label{f:geometry}
\end{figure*}

\section{The CALIFA mother sample}

In order to maintain flexibility in scheduling, the pool of galaxies available for observations in the CALIFA survey is somewhat larger 
than the expected number of total observations. This pool -- henceforth called the CALIFA `mother sample' (MS) -- is defined by the selection 
criteria detailed below. Galaxies are drawn from this pool for observation according to visibility alone, which should be close to random 
selection. At any given time, the set of actually observed CALIFA galaxies will therefore be a random subset of the MS. 
In the following we always refer to this MS when speaking about CALIFA galaxies, unless explicitly stated otherwise. 

\subsection{Selection of the Mother Sample}
\label{s:mother}

There were five main steps in the selection of the MS: 

\begin{enumerate}
\item \emph{Size selection:} The MS was selected from the SDSS DR7 \citep{abazajian09}. In CALIFA we are interested in nearby, bright galaxies. 
The SDSS spectroscopic sample suffers from incompleteness for objects brighter than an apparent magnitude of 14.5 in $r$. We 
therefore started with the PhotoObjAll catalogue of DR7 and selected objects that have 45\arcsec $< \isoA <$ 79.2\arcsec. Here 
$\isoA$ is the isophote major axis at 25 magnitudes per square arcsecond in the $r$ band\footnote{The exact meaning of all SDSSpipeline parameters is 
explained on the relevant DR7 webpage: \url{http://cas.sdss.org/astrodr7/en/help/browser/browser.asp}}. 
\item \emph{Quality assurance cuts:} We additionally applied cuts to avoid photometry problems as follows: a cut in Galactic latitude to exclude the Galactic plane: 
$b > 20^{\circ}$ or $b < -20^{\circ}$; a selection on a number of flags (NOPETRO = 0, MANYPETRO = 0, TOO\_FEW\_GOOD\_DETECTIONS = 0) 
to exclude obvious problems in the detections; a flux limit of $\mathit{petroMag}_{\mathrm{r}} < 20$ to exclude very faint objects. 
This yielded a sample of 1495 objects. 
\item \emph{Redshifts:} We then downloaded properties for all 1495 objects from SIMBAD\footnote{SIMBAD is a database that 
experiences frequent updates, such that 
the only way we can reference the `release' is by date -- January 15th 2010}. We used the `$\mathit{cz}$' redshifts when none 
were available from SDSS. For wavelength coverage reasons we restricted to redshift range to $0.005 < z < 0.03$. 
This discarded objects that were actually stars, but also those that had neither SIMBAD nor SDSS redshift. 
\item \emph{Visibility:} Finally, to reduce problems due to differential atmospheric refraction, it is best to keep the airmass $X < 1.5$.  The further limitation of 
hour angle to $-2$ h $< \mathit{HA} <$ 2 h (to cope with PMAS flexure problems) then limits the declination to $\delta > 7$ deg. 
Due to the sparsity of galaxies in the SDSS Southern area, this limit was only applied in the main SDSS area, i.e.~for Right Ascension 5h $< \alpha <$ 20h.
\item \emph{Final adjustments:} Of the nearly final sample of 942 objects, five were eliminated based on visual inspection 
(e.g. because they were part of a much larger galaxy that was shredded by the SDSS pipeline). 
Two objects were added later on by hand. One is 
NGC4676B, the second system of the Mice galaxies. This object was added because the other object in the pair falls in our MS. 
This gave us the opportunity to study a merger system and to relate its properties to the larger sample. 
Also, it would in principle fit our size criteria, if it had been treated properly by the SDSS pipeline. 
The other object, NGC5947, was observed due to a glitch with the database on the very first observing night. It however has properties very 
similar to objects in our main sample, so we left it in. To obtain a sample with the exact statistical properties described here, one 
would thus have to discard NGC4676B and NGC5947. 
\end{enumerate}

Within our final sky area there are only 18 objects which would have passed all our quality and size cuts but still have no redshift (942 have redshifts). 
Those objects are not part of the sample. This means that we are missing less than 2\% of our sample, even if all of these were 
at the right redshift. In the more likely case that their redshift distribution is similar to that of those galaxies with redshifts, 
we are missing 1.2\% of our sample. 

The final CALIFA MS that we describe in this paper thus contains 939 objects. The final observed sample 
will be a random sub-selection of the MS in all physical galaxy properties. Sub-selection happens according to visibility only. 
The sky and redshift distribution of the MS is shown in Figure \ref{f:geometry}. Note that absolute magnitudes in Figure 
\ref{f:geometry} are based on the analysis presented later in Section \ref{s:absmag}, which includes growth curve photometry of the 
CALIFA MS galaxies, hence the notation $M_{r,\mathrm{GC}}$. These absolute magnitudes have been corrected for foreground 
(Galactic) reddening, but not for internal attenuation. Absolute magnitudes based on SDSS 
Petrosian magnitudes and redshifts only will be used on the following as well for purposes of comparison to a bigger SDSS sample. 
For these we use the notation $M_{r,\mathrm{p}}$.

\subsection{Distances, spatial coverage of the IFU and linear scale}
\label{s:cover}

Distances for the MS have been obtained from NED and Hyperleda \citep{paturel03}. From NED we retrieved the distances as 
corrected for Virgo, Shapley and Great Attractor infall, \citep[][in which $H_0=71\,\mathrm{km\,s^{-1}\,Mpc^{-1}}$, 
which is so close to our fiducial value 
that we do not correct for the difference]{mould00}. We also retrieved redshift independent distances from NED. Hyperleda makes 
available distance moduli which are corrected for Virgo-centric infall and we also derived distances from pure Hubble flow for 
comparison. Unfortunately, redshift independent distances do not exist for all our galaxies. Also, they are inhomogeneous, 
sometimes significantly so. We therefore use them as a benchmark only. The best correlation with redshift independent 
distances was found for the NED-infall-corrected ones, which are available for all galaxies. 
We therefore adopted those as our fiducial distances. 

\begin{figure*}[tbp]
\begin{center}
\resizebox{0.49\hsize}{!}{\includegraphics[]{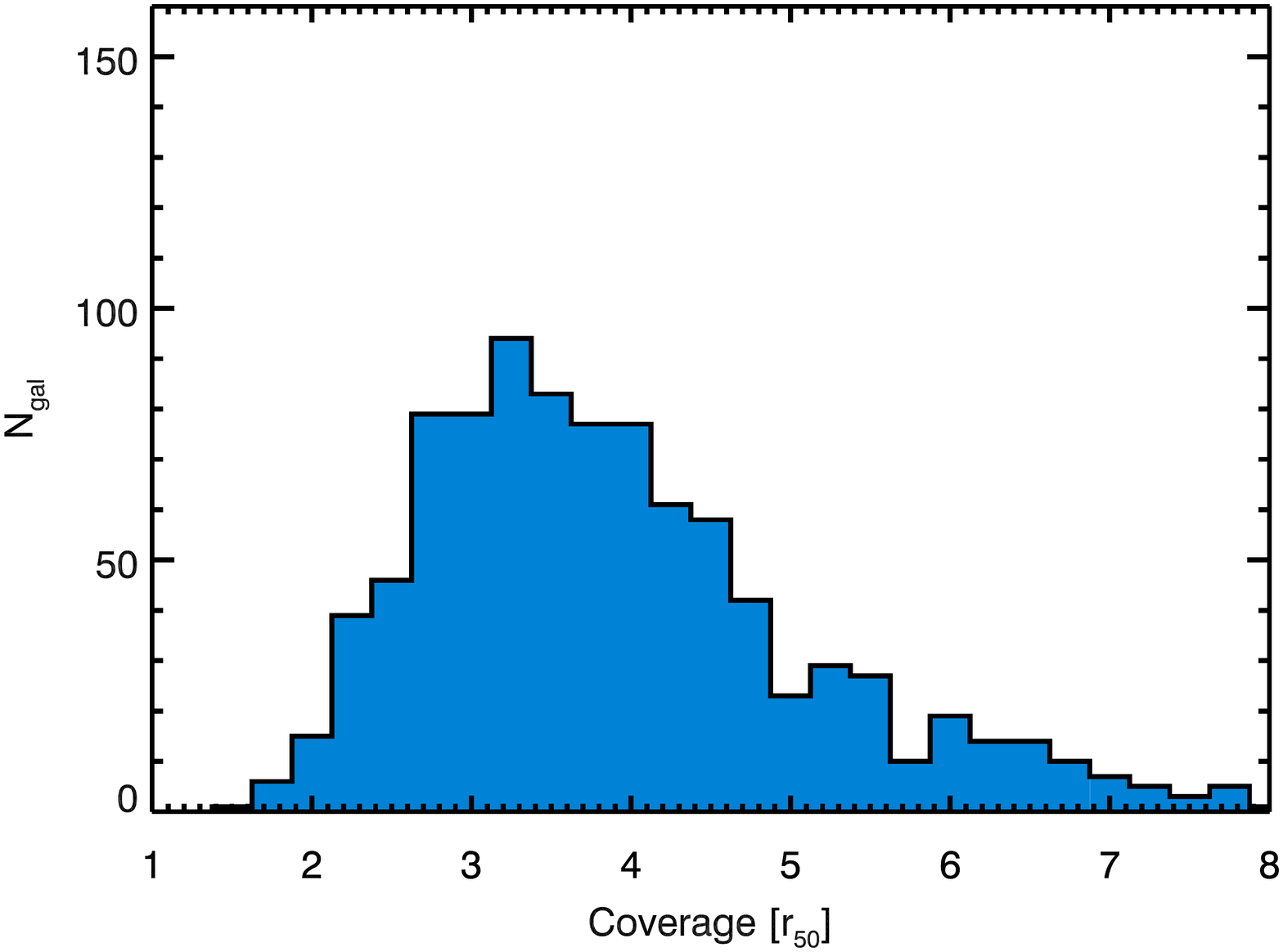}}
\resizebox{0.49\hsize}{!}{\includegraphics[]{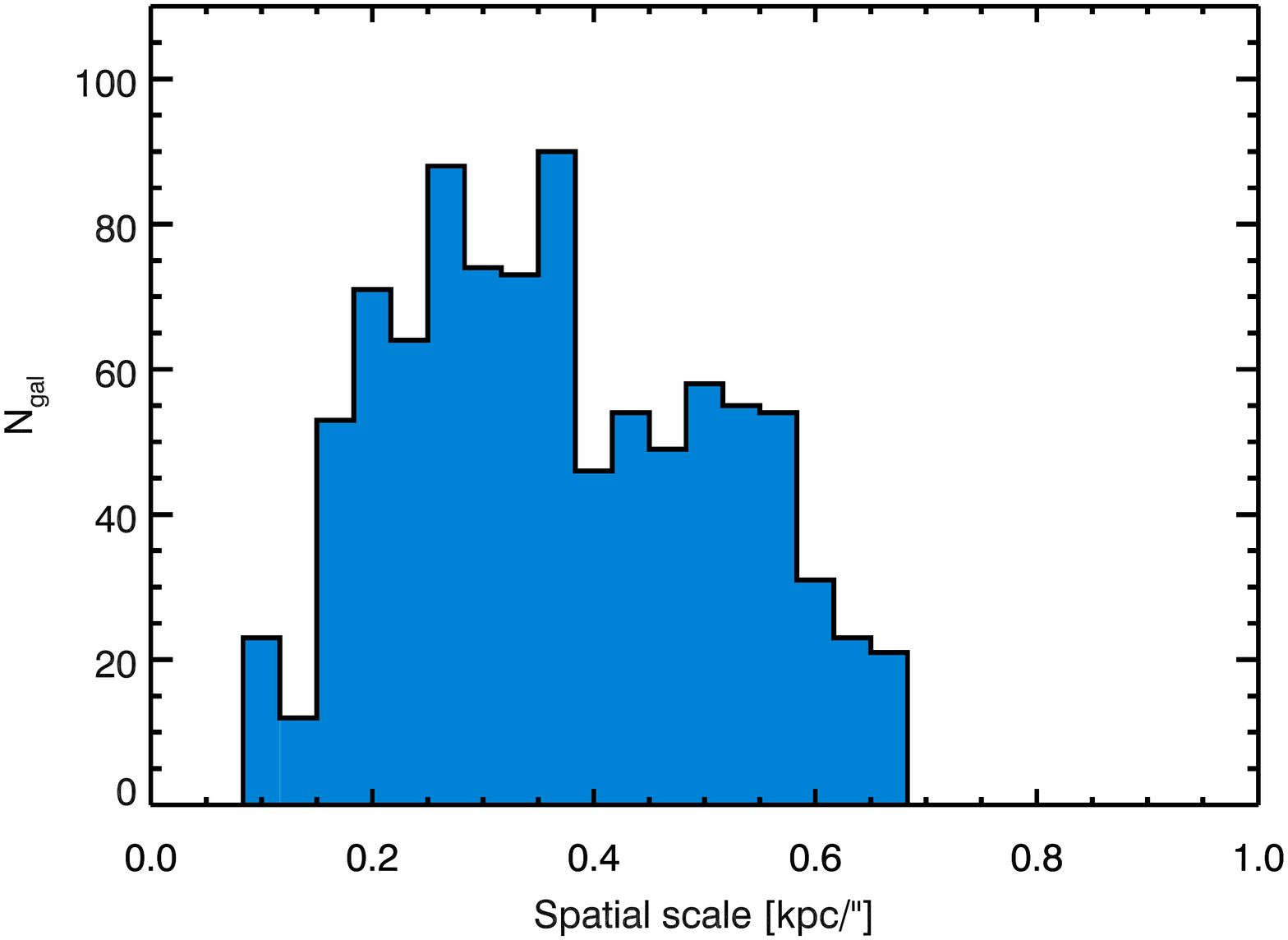}}
\end{center}
\caption[distances]{\emph{Left:} Histogram of radial coverages of the CALIFA galaxies, i.e.~ the ratio between the radius of the 
Field of View of PPak and $\mathit{petroR50}_{\mathrm{r}}$. This figure does not give the actual spectroscopic coverage, 
which may be smaller due to 
S/N issues. \emph{Right:} Histogram of spatial scales with which the CALIFA galaxies are observed. A fibre diameter is 2.7\arcsec, whereas 
the typical fibre-to-fibre distance is 3\arcsec. The final spatial resolution of CALIFA will depend on future optimizations of the 
cubing code, but will be approximately  3\arcsec. }
\label{f:distances}
\end{figure*}

CALIFA was designed to cover `galaxies over their entire optical extent' and it is useful to verify how much this is the case. 
Figure \ref{f:distances} therefore shows the histogram of radial coverage in units of the SDSS pipeline quantity 
$\mathit{petroR50}_{\mathrm{r}}$, called $r_{50}$ hereafter. Clearly, the overwhelming majority of our galaxies (97\%) are covered 
to more than $2 \times$ $r_{50}$\footnote{Note that this fraction drops to 50\% when using the more accurate $r_e$ 
from the growth curve analysis in Section \ref{s:growthcurve}, but the number given in the text above provides a 
natural comparison to other surveys.}. In most cases we indeed obtain useful data out to these 
large radii. The real depth of CALIFA data is described in detail in \citet{sanchez12a} and \citet{husemann13}. 
On average over the MS, the PPak IFU covers 1.4 times the isophotal diameter determined from the SDSS imaging, 
with the maximum and minimum values being 1.64 and 0.94, as per selection. 
 
Another useful number is the average spatial scale of the CALIFA data, also shown in Figure  \ref{f:distances}. 
The mean physical scale of one PPak fibre for the CALIFA MS is 1 kpc. The actual spatial resolution in the 
final data cubes delivered by CALIFA depends on the cube reconstruction software, which is still being optimized at the 
time of writing of this paper. Due to the three point dither pattern, we expect the final spatial resolution to be better 
than 1 kpc in the mean and better than 1.9 kpc for all galaxies in the CALIFA sample. CALIFA objects can thus not only 
be resolved in their different galaxy components (nucleus, bulge, disc, spiral arms), but due to the average distance between 
H\,{\small II} regions, even single H\,{\small II}  complexes can be identified and studied \citep[][]{sanchez12b}. Note  
that the redshift dependence of the size cuts in physical units and the intrinsic change of spatial resolution with redshift introduces 
a mass dependence of the spatial resolution as measured in kpc. This effect is approximately a factor of two between the highest and 
lowest redshift limits, but may still be important for some science applications.

\subsection{Multi-wavelength data available for the CALIFA sample}
\label{s:ancillary}

We have cross-correlated the positions of CALIFA galaxies with those in a variety of available databases covering many wavelength 
ranges. Table \ref{t:surveytable} indicates the number of CALIFA galaxies which have a match in each survey. Whether consistent 
integrated fluxes are available (yet) is another question. We derived optical fluxes for CALIFA galaxies from a 
growth curve analysis in Section \ref{s:growthcurve}. To obtain matched integrated fluxes from the other surveys by growth 
curve analysis would be prohibitive and not necessarily useful, due to the different depth and background characteristics. 
We therefore suggest to resort to either using catalogues that represent `total fluxes' as derived by these surveys, or to 
determine own fluxes based on the apertures defined by the isophotal position angle, axis ratio and half-light major axis
derived by the growth curve analysis.

\begin{table*}
\begin{minipage}[H]{\linewidth}
\caption{Available ancillary data}             
\label{t:surveytable}      
\centering 
\renewcommand{\footnoterule}{}         
\begin{tabular}{l c c c c}     
\hline\hline       
Survey/Telescope & Number of objects & Bands \\
\hline
SDSS & 939 & $u,g,r,i,z$ \\
2MASS  & 932 &  \it{J,H,K$_\mathrm{s}$} \\
IRAS & 243 & 12 $\mu$m, 25 $\mu$m, 60 $\mu$m,  100 $\mu$m  \\
WISE  & 939 &    W1,W2,W3,W4 \\
GALEX  & 655 &   FUV,NUV \\
HST  & 81 &  UV-NIR  \\
ROSAT  & 28 &   X \\
Chandra  & 42 &  X (u,s,m,h,b) \\
FIRST  & 814 &   1.4 GHz  \\
NVSS  & 939 &   1.4 GHz  \\
Spitzer  & 280 & 3.6 $\mu$m, 4.5 $\mu$m, 5.8 $\mu$m,  8 $\mu$m  \\
UKIDSS  & 267 &   \it{J,H,K,Y}  \\
\hline 
\hline                  
\end{tabular}
\end{minipage}
\end{table*}

The photometry used in Section \ref{s:absmag} was derived from the following resources:

{\bf 2MASS photometry: } 
The CALIFA MS table was cross-matched with the 2MASS All-Sky Extended Source Catalog (XSC) catalogue \citep{jarrett00}, 
providing $J$, $H$, $K_s$ photometry in Vega magnitudes. These were converted to AB magnitudes using offsets of 0.91, 1.39, 1.85, respectively 
\citep{blanton05a}. The CALIFA galaxy coordinates were used to find extended 2MASS source entries within 20\arcsec. For some galaxies 
the 2MASS coordinates can be significantly offset from the galaxy center by more than 10\arcsec. Such cases were deemed unreliable and 
were not used in the final match. 

{\bf GALEX photometry: } The CALIFA MS table was cross-matched with the GALEX GR6 database (using the GALEXView tool) 
for all GALEX `tiles' that have their centers within 0.55 degrees of a CALIFA galaxy. The magnitudes were determined from a growth 
curve analysis and should therefore be equivalent to the optical magnitudes. The photometry was computed following the recipes in 
\cite{gil-de-paz07}. The total number of galaxies observed is 663 and the total number of galaxies where we have 
useful photometry is 655. There are no FUV data for 52 of the 655 galaxies, either because the exposure time in the FUV is not 
sufficient, or because the galaxy is extremely red. More details on the UV photometry will be contained in a forthcoming 
paper (Catal\'{a}n-Torrecilla, in prep.).

\section{How CALIFA compares to the general galaxy population}
\label{s:represent}

The CALIFA survey was launched with the intention to characterize typical galaxies over a wide range of properties. 
This is in contrast to the samples of the SAURON project \citep{de-zeeuw02} and the ATLAS$^{3D}$ survey \citep{cappellari11}, 
which are focussed on early-type galaxies. Although some focussed projects using IFS on late type galaxies also exist 
\citep{ganda06, bershady10}, no existing survey using IFS has attempted to observe a sample of galaxies covering all types of galaxies.

We already demonstrated in \citet{sanchez12a} that the CALIFA MS covers the full area occupied by galaxies in the 
colour-magnitude diagram. In the following we address the issue of representation in a more rigorous way. We investigate which 
selection effects might be affecting the CALIFA sample, and we estimate the limits of representativity, outside of which the survey 
will not constrain the properties of galaxies in general.

\subsection{Comparison data}
\label{s:compsamp}

The current state of the art for low-redshift galaxy surveys is set by the spectroscopic part of the SDSS \citep{strauss02}, which has enabled 
extensive investigations of galaxy properties in the nearby Universe. It is therefore natural to compare the statistical properties of the CALIFA 
MS with those of much bigger and well-groomed SDSS galaxy samples. Note that since CALIFA is based entirely on the SDSS 
photometric database, any fundamental limitations in those data (such as the well-known bias against very low surface brightness galaxies) 
will translate directly into corresponding selection effects for CALIFA. We do not discuss such effects further, but refer the interested 
reader to \cite{kniazev04}.

Our comparison sample of galaxies extracted from the SDSS DR7 spectroscopic database is flux-limited to 
$\mathit{petroMag}_{\mathrm{r}} < 17.7$ and covers a geometric 
footprint in the sky of 8033 deg$^2$, very similar to (but slightly less than) the CALIFA footprint. We considered only galaxies with well-measured 
SDSS redshifts between $z = 0.003$ and $z = 0.1$; there are some 260\,000 galaxies matching this selection. For brevity, we denominate this 
set as `the SDSS sample' henceforth. For some of the tests presented below we further limited the outer redshift cut to the same value as for 
CALIFA, $z < 0.03$, which reduced the sample to 26\,900 galaxies; this we call `the low-$z$ SDSS subsample'. All relevant pipeline quantities 
such as apparent magnitudes, angular size estimates etc.\ are by construction consistent with those in the CALIFA tables, enabling direct 
comparisons. Note however that most of the SDSS galaxies are not only much fainter than the galaxies in CALIFA but also much smaller 
(in angular sizes), typically subtending no more than a few arcsec in the sky. This may lead to subtle systematic differences in some of the 
photometric quantities, due to the way the SDSS pipeline treats extended objects of different sizes, which ultimately limit the accuracy of 
this comparison.

\begin{figure}[tb]
\includegraphics[width=\hsize]{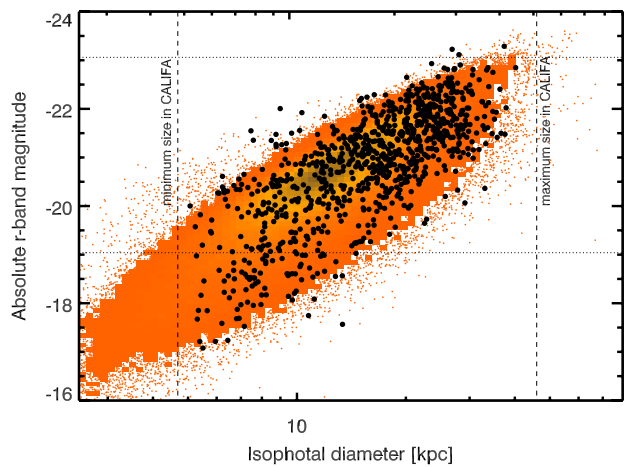}
\caption[Diso-Mr]{Selection limits of CALIFA: Absolute magnitudes $M_{r,\mathrm{p}}$ are plotted against linear isophotal sizes of galaxies 
in the CALIFA MS (black points), compared to the same for galaxies in SDSS (orange).
The two vertical dashed lines delineate the range of galaxies \emph{accessible} to CALIFA; all galaxies within this range would 
be selected by CALIFA if located at a suitable redshift. The horizontal lines represent the limits inside which for a certain luminosity 
bin the fraction of SDSS galaxies within the CALIFA `accessible range' is above 95\%.}
\label{f:Diso-Mr}
\end{figure}

\begin{figure}[tb]
~~\includegraphics[width=\hsize]{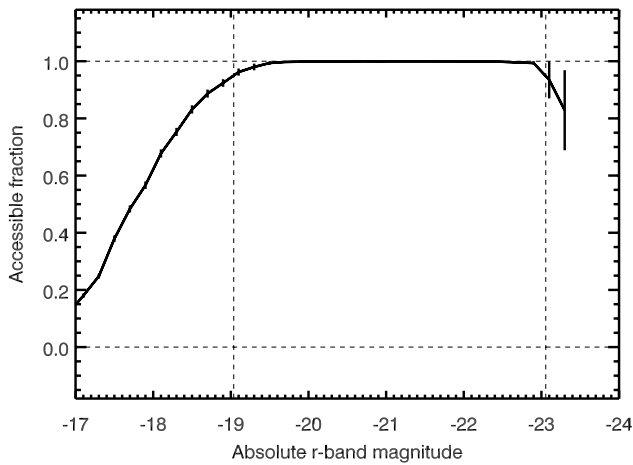}
\caption[selfuncMr]{Fraction of SDSS galaxies within the CALIFA accessible range of \Diso, as a function of absolute magnitude. 
Error bars are Poissonian. The two vertical lines bracket the range where the fraction is $>$ 95\%.}
\label{f:selfunc_Mr}
\end{figure}

\begin{figure}[tb]
\includegraphics[width=\hsize]{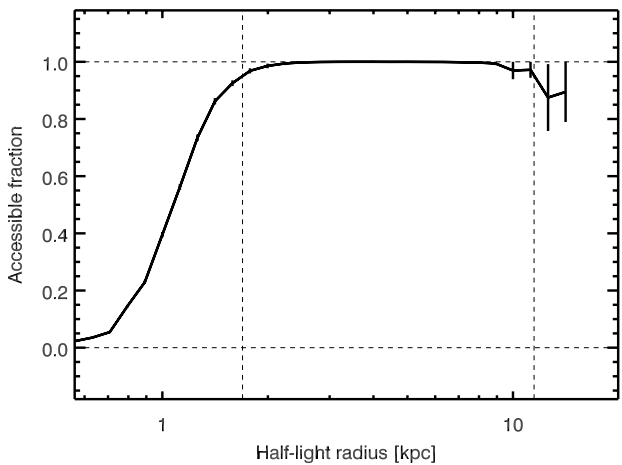}
\caption[selfunchlr]{As Figure \ref{f:selfunc_Mr}, but showing the fraction as a function of half-light radius (i.e. SDSS pipeline  $r_{50}$). 
The two vertical lines again bracket the range where the fraction is $>$ 95\%.}
\label{f:selfunc_hlr}
\end{figure}

\subsection{Limits of the CALIFA selection}
\label{s:selim}

We first investigate the question that users of public data from the CALIFA survey might find most relevant: What are the 
ranges in absolute magnitudes, stellar masses, and linear sizes (half-light radii) over which CALIFA provides a representative 
sample? How sudden or gradual is the transition when moving away from this range? And in particular, are there domains 
where CALIFA has a complicated selection function, for example where only the most compact or the most extended 
galaxies are included in the sample?

Under the assumption that the SDSS sample is a fair representation of galaxies in the local Universe, these questions can 
be empirically addressed by applying the CALIFA size selection criteria to SDSS galaxies. When doing this it is important 
to realize that whether or not a galaxy is in CALIFA depends \emph{only} on its linear isophotal size \Diso\ and on its redshift. 
While most SDSS galaxies have \emph{angular} sizes much too small for CALIFA, many of them have \Diso\ values that at 
some other (generally lower) redshift would make them accessible to the CALIFA criteria. Only galaxies with \Diso\ smaller 
than the smallest possible size $D\subsc{iso, min} = 4.7$~kpc -- corresponding to \isoA = 45\arcsec~at $z = 0.005$ -- 
would not make it into CALIFA at any redshift. Equally, the maximum possible linear size of any CALIFA galaxy corresponds to 
$79\farcs2$ at $z = 0.03$, or $D\subsc{iso, max} = 46$~kpc.

In Figure \ref{f:Diso-Mr} we plot absolute magnitudes $M_{r,\mathrm{p}}$ against linear sizes \Diso\ for both the SDSS and the CALIFA samples. 
For consistency between both samples, absolute magnitudes in this figure have been derived from the apparent Petrosian $r$ band 
magnitude as given by the SDSS photometric pipeline and distances have been calculated directly from the observed redshift (i.e.~ neglecting 
any corrections for peculiar velocities of the galaxies). 
All SDSS galaxies within the two vertical dashed lines, i.e.\ within the range 4.7~kpc $<$ \Diso\ $<$ 46~kpc, could and would be 
selected by CALIFA if located at a suitable redshift. For magnitudes $-19 \ga M_{r,\mathrm{p}} \ga -23$, essentially all SDSS galaxies are 
within this domain, irrespective of their actual sizes. We quantify this by marginalizing over \Diso\ and computing the fraction of 
SDSS galaxies within the CALIFA `accessible range'; this is shown in Figure \ref{f:selfunc_Mr} (with Poissonian error bars 
representing the number of SDSS galaxies in each bin). The fraction is above 95\% for the range
\begin{equation}
-19.0 > M_{r,\mathrm{p}} > -23.1
\end{equation}
and falls rapidly outside of that range. Notice that even the huge $z < 0.1$ SDSS sample contains only relatively few galaxies 
at $M_{r,\mathrm{p}} < -23$, so that the error bars are correspondingly large. 

Since \Diso\ is also correlated with half-light radius, we can perform the same exercise to determine the completeness with 
respect to that quantity. In Figure \ref{f:selfunc_hlr} we show the marginalised fraction of SDSS galaxies within the CALIFA 
accessible range of \Diso, now as a function of $r_{50}$. The `accessible fraction' is again higher than 95\% for the interval 
\begin{equation}
1.7 \:\text{kpc} < r_{50} < 11.5 \:\text{kpc} \:.
\end{equation}
We finally also estimated the corresponding limits in stellar masses, anticipating the results from Sect.~\ref{s:photometry}. 
We find that the fraction is above 95\% for the range 
\begin{equation}
9.65 < \log (M^\star/\Msun) < 11.44\:.
\end{equation}

Only outside of these `completeness limits' does the CALIFA selection function depend on galaxy size in a non-trivial way, in 
the sense that low-luminosity galaxies can get into CALIFA only if they have a large value of \Diso (see also Sect.~\ref{s:axrat}), 
and very high-luminosity galaxies may be captured in CALIFA only if they are abnormally small. However, less than 10\% of all 
galaxies in the CALIFA MS are located in these `outside' regions of parameter space, most of them forming the low-luminosity 
and low-mass tail of the sample. For statistical purposes they should be left out of consideration.

Of course, only very few of the SDSS galaxies actually made it into the CALIFA sample; most are at too high redshifts and 
appear therefore as too small. But as long as the isophotal size distribution function is the same everywhere, this selection 
can be accurately quantified in terms of the formal survey volumes for CALIFA and SDSS, which we discuss in the next 
subsection. We thus conclude that for the given range in luminosities and masses, the apparent diameter selection 
does not introduce any size bias.

\begin{figure}[tb]
\includegraphics[width=\hsize]{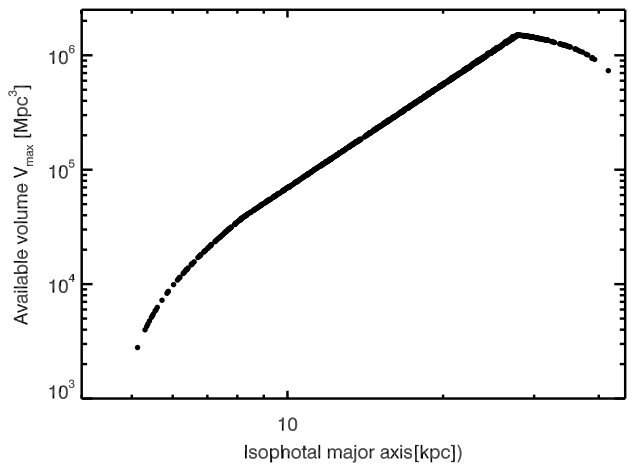}
\caption[vmax]{Available survey volume for all galaxies in the CALIFA MS, as a function of linear isophotal size 
as derived from the observed redshift.}
\label{f:vmax}
\end{figure}

\section{Volume corrections and galaxy number density distributions}
\label{s:volcor}

\subsection{CALIFA survey volume}
\label{s:volume}

The CALIFA footprint on the sky subtends $\Omega$\subsc{C} = 8700 deg$^2$, see also Figure \ref{f:geometry}. Together with the sample 
redshift range of $0.005 < z < 0.03$, this converts into a formal comoving volume of $\sim 1.7 \times 10^6$ Mpc$^3$ (adopting 
the cosmological parameters specified in Sect.~\ref{s:intro}). This, however, is not the actually available volume for the galaxies 
in the survey: Because of the narrow range in permitted angular sizes (less than a factor of 2), any galaxy of given linear size 
is included in the CALIFA selection only over an object-dependent range in redshifts (see Figure \ref{f:geometry}). 

The available volume \emph{per galaxy} can be computed with the \Vmax\ method by \citet{schmidt68}, the application of which 
is straightforward for a diameter-limited sample \citep[e.g.,][]{de-jong94}. For CALIFA we assumed that the ratio between 
apparent and linear isophotal size of a galaxy depends only on its angular diameter distance (i.e.\ we neglected cosmological 
surface brightness dimming, and any `K correction in size'). We furthermore assumed pure Hubble flow distances, 
which should be a good approximation for most objects in the sample but may introduce distance errors of up to $\sim 20$\% 
for the lowest redshift galaxies. We then computed, for each galaxy in turn, the minimum and maximum redshifts for which an 
object of the same linear size \Diso\ would still be captured by the CALIFA selection criteria. The available volume \Vmax\ follows 
directly from these object-specific redshift limits and the survey solid angle. It is easy to see that \Vmax\ depends \emph{only} 
on the value of \Diso\ of a galaxy. Figure~\ref{f:vmax} shows the variation of \Vmax\ with \Diso\ for the CALIFA MS. The maximum 
volume of $1.5\times 10^6$ Mpc$^3$ is reached for big galaxies located somewhat below the outer redshift boundary, 
while smaller (and therefore less luminous but more numerous) galaxies have much lower \Vmax\ values.

These numbers are applicable to the full MS. At any given time, only a fraction $f_{\mathrm{gal}} < 1$ of all galaxies in that sample 
will have IFU data. Assuming that the observed objects constitute a random subset of the MS, this reduction can be condensed 
into an `effective solid angle' $\Omega\subsc{eff} = f \times \Omega\subsc{C}$, and thus the value of $\Omega$\ computed for the mother 
sample simply has to be corrected by the same factor $f_{\mathrm{gal}}$, which again translates into correcting downwards 
the \Vmax\ values of each galaxy downwards by the same factor. 

\begin{figure}[tb]
\includegraphics[width=\hsize]{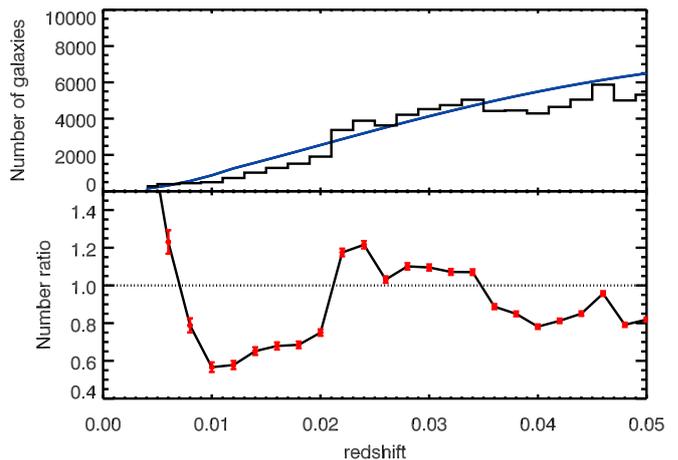}
\caption[]{Top: Observed (black) and predicted (blue) number of SDSS galaxies with magnitudes 
$r<17.7$ per $\Delta z = 0.002$ redshift 
bin. Bottom: Ratio of these two numbers, as a function of redshift.}
\label{f:rdg}
\end{figure}

Before turning to apply these volume corrections to the CALIFA sample we have to take another effect into account, namely 
variations in the galaxy number density due to large-scale structure. These variations are significant even when averaging 
over $\sim 10^6$~Mpc$^3$. We obtained a quantitative estimate of the magnitude of the effect on the CALIFA survey volume by 
the following procedure: We subdivided the SDSS comparison sample into redshift bins of $\Delta z = 0.002$ and counted 
the number of galaxies per bin. We then calculated, in each bin, the total number of galaxies \emph{expected} from the 
Schechter fit to the `local cosmic mean' luminosity function by \citet{blanton03b}, taking into account the apparent 
magnitude limit of the SDSS spectroscopic sample and `evolving' the luminosity function from $z=0.1$ to the mean redshift of 
each bin. The ratio of these two numbers provides an estimate for the redshift-dependent deviation of the number density of 
galaxies from the cosmic mean, averaged over scales corresponding to $\Delta z = 0.002$ and the SDSS DR7 footprint. The 
result is displayed in Figure \ref{f:rdg}, showing that the variations amount to more than a factor of 2 between minimum and 
maximum redshift, for the CALIFA redshift range of $z<0.03$. We note that a conceptually similar plot was already shown by 
\citet{blanton05b} only for the much smaller DR2 footprint and using infall-corrected redshift distances rather than plain redshifts. 

We can now use these ratios to apply redshift-dependent correction factors to the galaxy number density. Doing so however 
implies a number of simplifying assumptions: (1) We neglect the differences in footprints between SDSS-DR7 (spectroscopic sample) 
and CALIFA. (2) We consider only variations as a function of redshift and neglect transverse effects. (3) We assume that the 
\emph{shape} of the LF is always the same, only the normalization varies. Applying the correction is simple: If at the redshift $z$ 
of galaxy X the relative under- or overdensity is $\delta(z)$, we give galaxy X a weight $1/\delta$. Mathematically this is equivalent 
to combining the inverse volume $V_\mathrm{max}$ and the density factor $\delta$ into a single volume weight 
$V'_\mathrm{max} = \delta\times V_\mathrm{max}$ and then use $V'_\mathrm{max}$ for all volume corrections. We demonstrate 
the relevance of this correction in the next subsection. 

We thus conclude that the CALIFA MS is a statistically well-defined subset of the local galaxy population, with easily 
computable and quite accurately known volume weight factors per galaxy. It is important to keep in mind that any mean values 
computed directly from the observed sample (i.e. not corrected for survey volume) will be different from those of any other sample. 
In the next subsection we use these weights to explore how well CALIFA represents the mix of different galaxy types in the 
local Universe.

\begin{figure}[tb]
\includegraphics[width=\hsize]{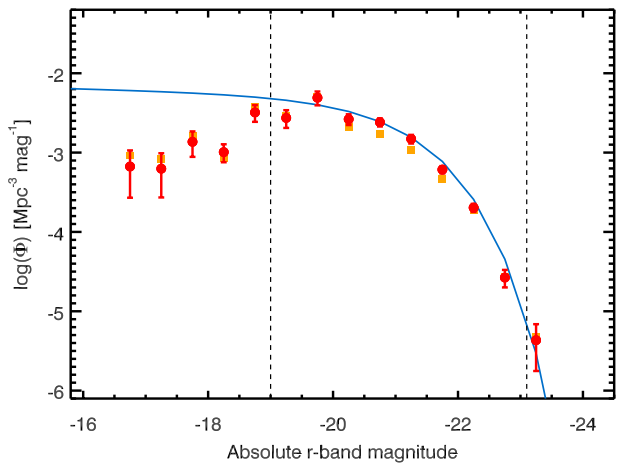}
\caption[CALIFA_LF]{ The red points show the $r$ band luminosity function of galaxies estimated from the CALIFA MS, using 
absolute magnitudes from the SDSS $M_{r,\mathrm{p}}$ and 
with error bars representing Poissonian uncertainties only. The orange squares are for the same sample, but without the corrections 
for variations in cosmic density. The blue solid line shows the Schechter fit to the LF of \citet{blanton03b}, 
adjusted to our cosmology and redshift range. The vertical dashed lines indicate the completeness limits derived in Section 
\ref{s:selim}. The faintest magnitude at which the luminosity function itself is marginally consistent with that of Blanton et al. would 
imply that the limit of completeness for the CALIFA sample is at roughly $M_{r,\mathrm{p}} < -18.6$. } 
\label{f:CALIFA_LF}
\end{figure}

\subsection{Luminosity function}
\label{s:lumfun}

We now investigate whether the overall number density of galaxies estimated from the CALIFA diameter-selected sample is 
in line with expectations from other surveys, thus whether or not CALIFA might be missing a significant fraction of galaxies. 
We also consider if galaxies of different luminosities are represented in adequate proportions by the sample. 

To this purpose we constructed the binned $r$ band luminosity function (LF) from the CALIFA MS using the 
\Vmax\ estimator and compared it with the LF estimated from SDSS. While there are more sophisticated methods 
available for measuring luminosity functions, we are mainly interested in a global comparison for which the simple \Vmax\ 
approach is sufficient. For the same reason we also did not attempt to apply any corrections for photometric incompleteness 
which would affect SDSS and CALIFA equally. We computed space densities both with and without the redshift-dependent 
correction for large-scale structure derived in the previous subsection. We did not apply k-corrections for this exercise, as these 
are very small for the redshift range considered. 

For comparison we again used the Schechter function fit to the LF constructed from almost 150\,000 SDSS galaxies by 
\citet{blanton03b}, adjusted to our cosmology and `evolving' the LF from $z = 0.1$ to the mean redshift of the CALIFA sample. 
The outcome of this comparison is shown in Figure \ref{f:CALIFA_LF}, demonstrating that CALIFA allows us to estimate the 
galaxy number density and luminosity function for absolute magnitudes $M_{r,\mathrm{p}} < -18.6$ with reasonable fidelity.

While the LF computed from the CALIFA sample without density correction (shown as orange squares in 
Figure \ref{f:CALIFA_LF}) is already quite close to the one from SDSS, 
the differences in some points are certainly greater than the Poissonian error bars. However, an accurate match 
would be purely fortuitous given the significant redshift-dependent modulations in galaxy number density shown in 
Figure \ref{f:rdg}. But when we apply the redshift-dependent correction (i.e. using the effective volume weights 
$V'_\mathrm{max}$ defined above), the agreement becomes almost perfect. Recall that while the correction is 
\emph{applied} to the CALIFA sample, it was \emph{derived} from the full SDSS sample alone without any reference to CALIFA. 
It is remarkable that both the overall normalization and the relative distribution of luminosities are captured so well by 
the CALIFA sample, given that it comprises less than 1000 galaxies.

At luminosities below $M_{r,\mathrm{p}} \approx -18.6$, the LF from CALIFA turns over and stays below the SDSS LF. This indicates the 
expected incompleteness at low luminosities, which in turn is a direct consequence of the low-redshift limit of CALIFA that 
excludes dwarf galaxies with \Diso$ < 4.6$~kpc. While there is also a related high-luminosity completeness limit at 
$M_{r,\mathrm{p,min}} = -23.1$ due to the upper redshift cut, this limit is actually washed out by small number statistics: 
According to the luminosity function, the number density of galaxies at  $M_{r,\mathrm{p}} = -23$ is approximately $10^{-6}$~Mpc$^{-3}$, 
which in combination with the maximum survey volume (Figure \ref{f:vmax}) implies that the total number of such galaxies 
expected for CALIFA is of order unity. In other words, galaxies more luminous than $M_{r,\mathrm{p}} = -23$ might be missing if 
they are too extended, but already independently of size they are largely absent in CALIFA because the survey volume 
is too small. 

These comparisons demonstrate that in terms of total number density and the distribution of luminosities, the CALIFA 
MS is very close to a fair representation of non-dwarf galaxies in the local Universe. 

\begin{figure}[tb]
\includegraphics[width=\hsize]{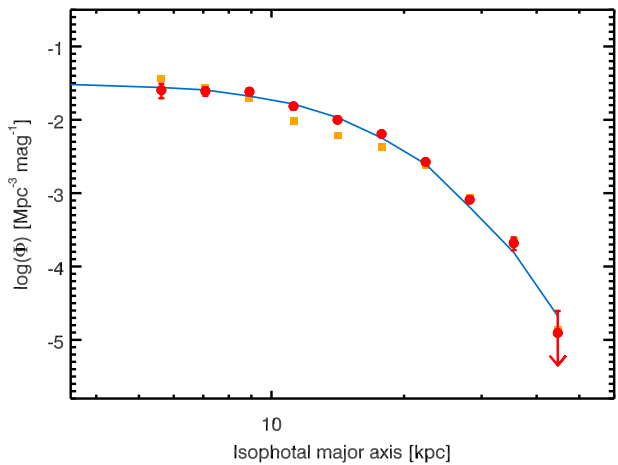}
\caption[CALIFA_SDF]{The distribution function of linear isophotal sizes \Diso\ of galaxies estimated from the CALIFA 
sample, compared to the same distribution constructed by us from the SDSS low-$z$ subsample. Symbols and line 
types as in Figure \ref{f:CALIFA_LF}. } 
\label{f:CALIFA_SDF}
\end{figure}

\subsection{Size distribution function}
\label{s:SDF}

A distribution related to the LF is the size distribution function (SDF), quantifying the differential number density of 
galaxies at a given linear size. We use here the isophotal sizes \Diso\ and construct a binned estimate of the SDF 
in the same way as the LF. The result is depicted in Figure \ref{f:CALIFA_SDF}, again with redshift-dependent 
number density correction, together with the SDF determined by us from the SDSS low-$z$ subsample. Notice that the 
number density $\phi$ is given here per logarithmic decade. 

The agreement is again satisfactory, especially after density correction. This plot also shows (more clearly than in the 
LF) that CALIFA as a sample covering only a small range of apparent sizes reacts differently to large scale structure 
than a survey with a one-sided flux-limit. Consider the CALIFA points at $\log(\Diso/\mathrm{kpc}) \sim 1.1$. When uncorrected, these points 
deviate most strongly from the SDSS-based SDF. Figure~\ref{f:geometry} shows that galaxies with these sizes in 
CALIFA are located at redshifts around and just below $z \approx 0.015$, where the underdensity in the local Universe 
happens to be most pronounced (see Figure \ref{f:rdg}). Galaxies located there will be too rare in the sample compared to the cosmic mean. 
Thus, large-scale structure affects the shape of the resulting distribution function from CALIFA, whereas for a sample 
with a one-sided flux limit it mainly modulates the overall normalization. In both cases it is of course possible to correct 
for such effects, provided that the variations as a function of redshift are known.

\begin{figure}[tb]
\includegraphics[width=\hsize]{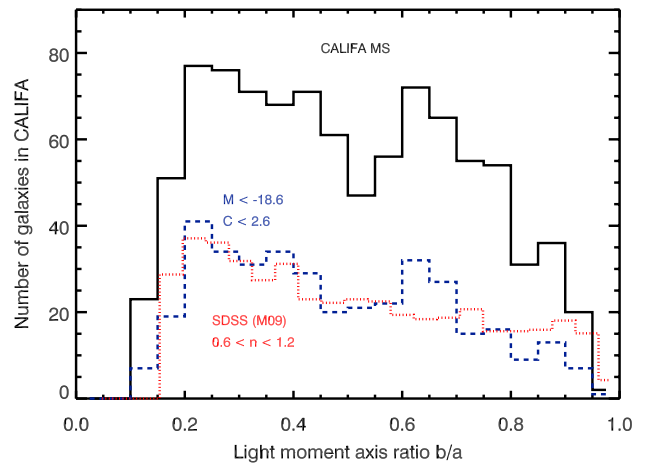}
\caption[]{Histogram of axis ratios (2nd order moments of the $r$ band light distribution) for the CALIFA MS. 
Overplotted in blue is the histogram for disc-dominated systems with $M_{r,\mathrm{p}} < -18.6$ and concentration indices $c < 2.6$, 
and in red for comparison the axis ratio distribution (rescaled to the same number of objects) for the disc-dominated galaxies in 
the SDSS sample of \citet{maller09}.} 
\label{f:axrat1}
\end{figure}

\begin{figure}[tb]
\includegraphics[width=\hsize]{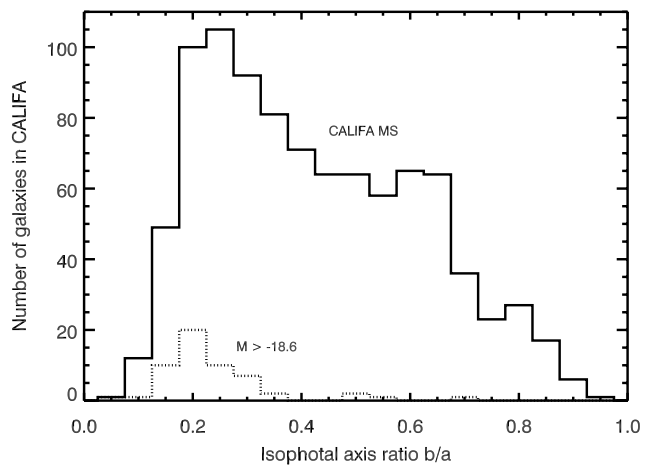}
\caption[]{Histogram of isophotal axis ratios (at 25 mag/arcsec$^2$ level) for the CALIFA MS. Overplotted with a dotted line  
is the histogram for the 55 low-luminosity systems with $M_{r,\mathrm{p}} > -18.6$, which are almost all highly inclined disc systems. } 
\label{f:axrat2}
\end{figure}

\begin{figure}[tb]
\includegraphics[width=\hsize]{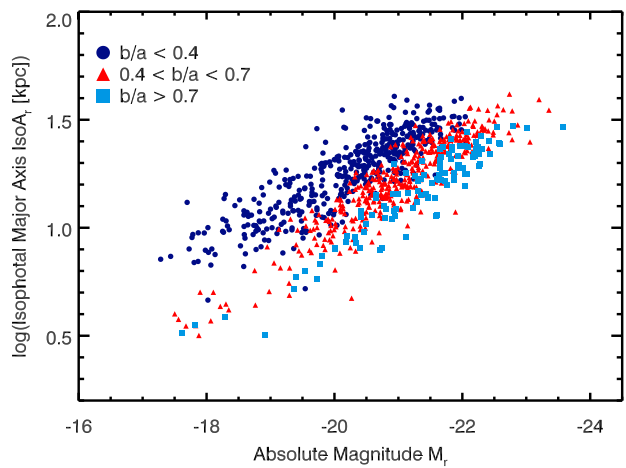}
\caption[]{Relation between absolute magnitudes ($M_{r,\mathrm{p}}$, uncorrected for internal extinction) and linear isophotal 
sizes of the CALIFA MS, colour-coded according to isophotal axis ratios.  } 
\label{f:axrat3}
\end{figure}

\section{The faint limit of the sample and the axis ratio distribution}
\label{s:axrat}

We now come to a property where we expect noticeable selection effects. It is long known that isophotal sizes 
of flattened, transparent (no attenuation) galaxies vary with inclination, simply due to the  projected change of surface brightness \cite[e.g.,][]{opik23}. 
It is therefore easier for an inclined disc galaxy to get into a sample defined by a minimum apparent isophotal size than it is for 
a face-on system of the same intrinsic dimensions. The magnitude of this effect depends on the degree of transparency; it is 
strongest for a fully transparent galaxy, and it disappears when the system is opaque, so that only its surface is observed. 
Notice that exactly the opposite selection effect exists for flux-limited surveys, favouring face-on systems over inclined ones. 
In this case the effect is significant when extinction is large, while it is negligible for transparent galaxies.

Yet, inclination is not an easily measurable quantity. For highly flattened (disc-dominated) systems the ratio between minor 
and major photometric axes can be used as a proxy. We thus expect that the CALIFA sample might display an excess of 
galaxies with low axis ratios, at least among disc-dominated systems, compared to a volume-limited sample. Such a dataset 
was constructed based on the SDSS by \citet[][hereafter M09]{maller09}, with the explicit purpose to statistically constrain 
the intrinsic shapes of galaxies. Axis ratios from the 2nd order moments of the light distribution were obtained in Section 
\ref{s:photometry}. Moment based axis ratios give similar results as those obtained from fitting S\'ersic models to the surface 
brightness distribution of galaxies, thus they provide a fair comparison to the results of M09. 

In Figure \ref{f:axrat1} we show the overall histogram of light-weighted axis ratios of the CALIFA MS, which turns out to be 
almost flat. Since any inclination-dependent selection effects should be most clearly seen for intrinsically flat disc-dominated 
galaxies, we separated the MS into early and late types by their concentration indices $c \equiv r_{90}/r_{50}$ in the $r$ band, with 
the dividing value at $c = 2.6$ \citep[e.g.,][]{strateva01,lackner12}. Figure \ref{f:axrat1} also shows the axis ratio distribution 
of only the $c < 2.6$ (= disc-dominated) galaxies, additionally limited to absolute magnitudes $M_{r,\mathrm{p}}$ brighter than $-18.6$ 
(cf.\ Sect.~\ref{s:selim}). For comparison the corresponding distribution of low S\'ersic index ($n < 1.2$) galaxies from the 
approximately volume-limited sample of M09 is also plotted, rescaled to match the corresponding number in the 
CALIFA sample. These two histograms are apparently very similar, indicating that the 
selection method for CALIFA does \emph{not} strongly bias the axis ratio distribution of luminous disc galaxies in the sample. 

We caution that light moment based axis ratios are weighted by light and thus are more sensitive to the brightest parts of a galaxy. 
Especially in presence of bulges, they tend to underestimate the axial ratio of the disc component, which is instead well represented 
by the outer isophotes. We therefore also considered the alternative approach of using the SDSS photometric pipeline delivered 
isophotal major and minor axes (\isoA~and \isoB) that can be combined into an axis ratio at the outer 25 mag/arcsec$^2$ level. 
The histogram of \emph{isophotal} axis ratios in Figure \ref{f:axrat2} is now clearly skewed towards low values of $b/a$, 
providing \emph{some} indication for the above selection effect in the CALIFA sample. To understand this better, we show as 
dotted histogram (in red) in Figure \ref{f:axrat2} the 55 galaxies of the CALIFA MS with $M_{\mathrm{r}} > -18.6$, thus below the 
completeness limit. Nearly all of these have axis ratios below 0.4 (this remains true when taking light-weighted axis ratios instead); 
visual inspection of the images confirms that these are predominantly disc-dominated systems seen close to edge-on. Presumably 
very few of these galaxies (if any) would have made it into the CALIFA sample if seen face-on; their angular sizes have been 
boosted through inclination, just enough to elevate them into the sample. We note in passing that in our flux-limited SDSS comparison 
sample we can directly verify the opposite trend mentioned above, namely that the distribution of isophotal axis ratios is skewed 
towards larger values. This is a direct consequence of non-negligible extinction in the $r$ band acting on highly inclined 
systems \citep[e.g.~][]{disney89,boselli94,unterborn08,padilla08}.

Figure \ref{f:axrat3} shows how inclination increases the isophotal sizes and weakens the magnitudes of disc-dominated galaxies, 
leading to a widening of the apparent luminosity-size relation. Take two galaxies with the same intrinsic size and luminosity, one seen 
face-on, one edge-on. While the face-on galaxy will be seen at its original position within the size-luminosity relation, the one seen 
edge-on will be shifted towards fainter magnitudes and larger isophotal major axis, i.e. perpendicular to the size-luminosity 
relation itself. The exact mix of internal extinction and surface brightness boosting due to inclination will depend on the galaxy 
type and thus presumably also on luminosity and mass. We make no attempt here to disentangle the two effects.

While the CALIFA sample thus has a higher proportion of inclined disc galaxies \emph{at the faint end}, the overall effect 
is not large. When using a light-weighted estimate of axis ratios, there is in fact no significant difference to the volume-limited 
sample of M09; when adopting axis ratios measured at an outer isophote the effect becomes more noticeable. Specifically for 
the galaxies close to and below the low-luminosity completeness limit there is at any rate a clear surplus of galaxies with very 
high inclinations in the CALIFA sample. 

We finally note that the ability to perform volume corrections for the CALIFA sample is completely unaffected by this possible 
selection bias for inclined galaxies. For any given galaxy, the available volume \Vmax\ depends only on its \emph{observed} 
size and on its redshift; moving a galaxy in- or outwards until it leaves the sample selection corridor has obviously no 
consequence for its inclination.

\section{Photometry, morphology, and stellar masses}
\label{s:photometry}

The SDSS pipeline has been optimized for a large survey and it was clear from the outset that the catalogued photometric 
properties for our sample would have to be verified. In particular, the CALIFA MS galaxies are bigger on the sky than the objects the 
SDSS pipeline has been optimized for. The SDSS pipeline Petrosian fluxes for the CALIFA MS therefore are likely to be affected 
in a different way than for a typical, large SDSS sample in the sense that their fluxes will be biased even lower as compared to the 
usual offset between the likely total flux from the galaxy and the Petrosian flux. We therefore set out to produce photometric quantities 
attempting to sum up all the available flux per galaxy using our own analysis. The reader should bear in mind that biases in comparisons 
between different samples will arise if the techniques used to obtain the photometry differ strongly. 

\subsection{Growth curve analysis}
\label{s:growthcurve}

The first step to obtain reliable integrated photometry from the images was to produce growth curve (GC) photometry for all sample galaxies 
in all bands. We used images from DR7. We first constructed masks for bright stars and background galaxies. In a first pass, 
masks were produced from the segmentation image of SExtractor \citep{bertin96}. These were then extended by hand for the regions 
within the galaxies, as SExtractor is not able to reliably identify foreground objects within galaxies. 
Neglecting the flux from masked regions would have led to 
systematic underestimation of galaxy brightness. In order to evaluate and include the missing flux from masked areas, we interpolated 
the masked regions using an inverse-distance weighted average. 
In order to apply the masks (corresponding to $r$ band images) to all 5 SDSS bands, we measured the shift between the different 
images and their $r$-band counterparts using their WCS (FITS World Coordinate System) RA and Dec coordinates, then shifted and 
cropped the masks accordingly. Inspecting the masked images visually, one sees that some light still spills out from rectangular 
masked regions, and some faint stars are left unmasked as well. While this would mean that the `real' sky flux is systematically 
overestimated, our galaxies are extended so it is likely that they also contain such unmasked foreground objects. 

The position angle ($\mathit{PA}_{\mathrm{gc}}$) and axis ratio ($\mathit{b/\!a}_{\mathrm{gc}}$) values were obtained by 
calculating light moments \citep[see Section 10.1.5 of the SExtractor 
manual vs2.13 and][]{bertin96}. The final $\mathit{b/\!a}_{\mathrm{gc}}$ value is the mean of the axis ratios of ellipses containing 
50\% and 90\% of the total flux. This is motivated as a compromise between a correct representation for most of the light (and thus 
correct derivation of the half light major axis) and a correct representation of the galaxy outskirts (and thus correct derivation of 
the total light). 

To derive the actual growth curve, all pixels on ellipses with successively incremented major axes and with fixed 
$\mathit{b/\!a}$ and $\mathit{PA}$ were summed up. 
If we were fitting the flux profile in sufficiently wide rings using simple linear regression, the best fit line should become 
horizontal at some radius, which we might then consider to be the edge of the galaxy. This statement would assume that galaxy flux
falls off asymptotically until it is indistinguishable from the sky fluctuations. In practice this is not the case, given that incomplete 
masks, light from other objects and sky gradients make the best fit slope switch from negative to slightly positive at some point. 
We opted for a solution in which we fit 150 pixel wide sections of the flux profile using simple linear regression, with neighbouring 
fit sections overlapping  by 100 pixels. When the flux profile slope becomes non-negative, we take the mean of the current ring 
as the sky value, and the ellipse with major axis value at the middle of the ring as the galaxy's edge.
We have verified that this procedure gives good results and is robust even in the presence of masked regions or faint unmasked 
objects. We added simulated de Vaucouleurs and exponential profiles to real sky backgrounds, including those with various defects, 
and ran the growth curve code on them. The procedure recovers practically 100\% of the flux for both de Vaucouleurs and exponential 
profiles.

The determined sky is of course very important for extended objects such as our sample galaxies. We thus subtracted the 
sky from the images before constructing the growth curve. We verified that there 
are no significant differences between the sky measurements from the SDSS pipeline and from the growth curve routine. 

The growth curve procedure was repeated with circular apertures for comparison purposes. The half-light major axes 
($\mathit{HLMA}$, for elliptical apertures) and half-light radii ($\mathit{HLR}$, for circular apertures) were calculated 
once the total extent and flux of a galaxy were known. We use the difference between circular and elliptical growth curve 
magnitudes as an indication of the uncertainty on each magnitude. The standard deviation of this scatter is 0.14 mag. 
We find that the resulting magnitudes are indistinguishable in a systematic way. The same is not true for the $\mathit{HLR}$, which is highly dependent on the projected inclination. 
The $\mathit{HLMA}$  depends less on inclination and we therefore consider it to be a better measurement of the true 
half-light radius of galaxies. We henceforth denote the $\mathit{HLMA}$  as $r_{\mathrm{e}}$ to distinguish it from the 
$r_{50}$ based on SDSS pipeline Petrosian fluxes. We will adopt growth curve measurements based on the elliptical 
annuli from here on. 

\subsection{Comparison of photometric measurements}
\label{s:photcompare}

It is instructive to compare the photometric measurements made in this section with the SDSS DR7 pipeline values as well as 
the values from the RC3 catalogue \citep{de-vaucouleurs91,corwin94}. While our measurements are based on the same data as 
the DR7 pipeline values, they are more comparable to the RC3 values in terms of the method used to recover them. 

\begin{figure*}[tbp]
\begin{center}
\resizebox{0.49\hsize}{!}{\includegraphics[]{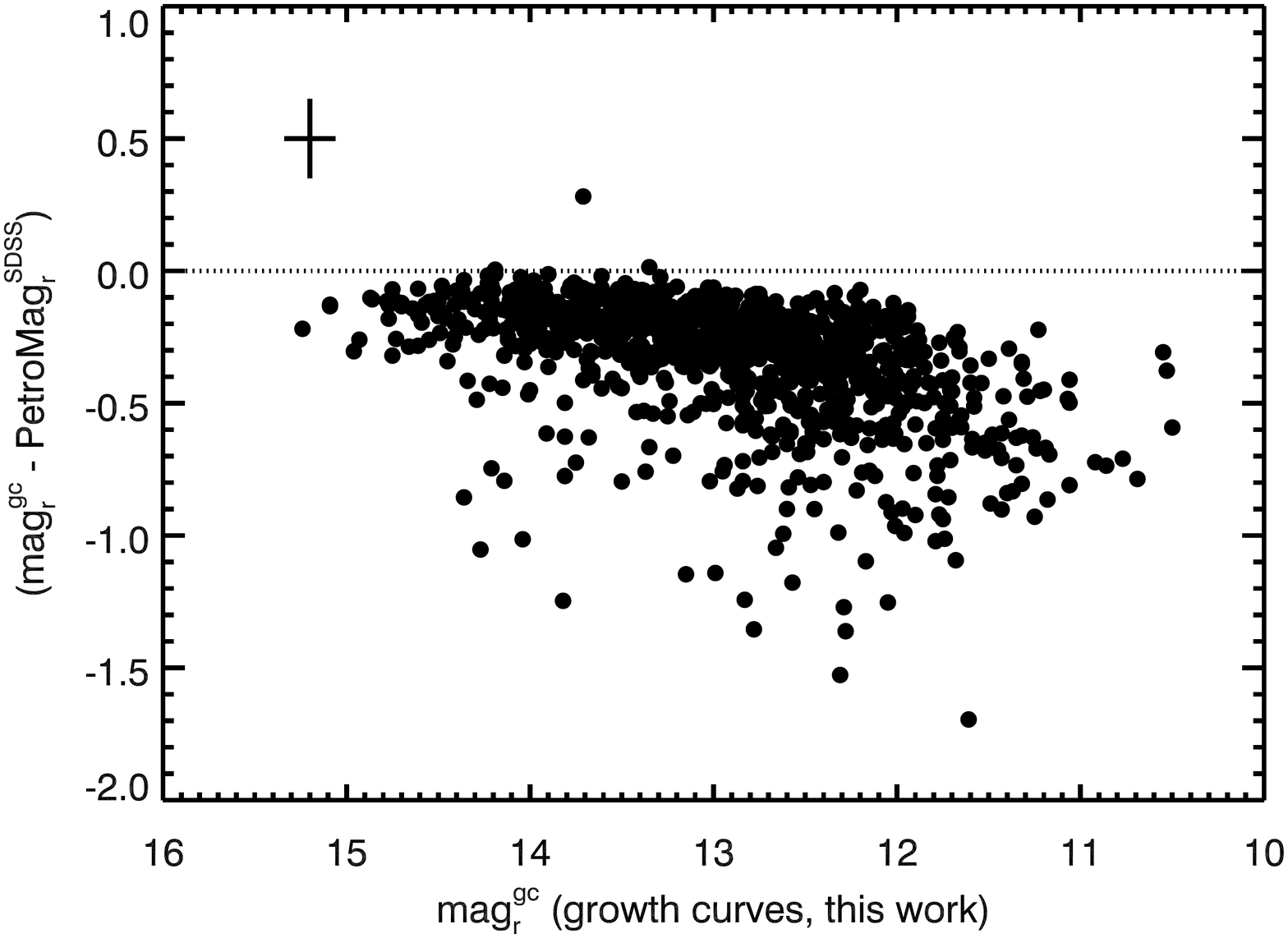}}
\resizebox{0.49\hsize}{!}{\includegraphics[]{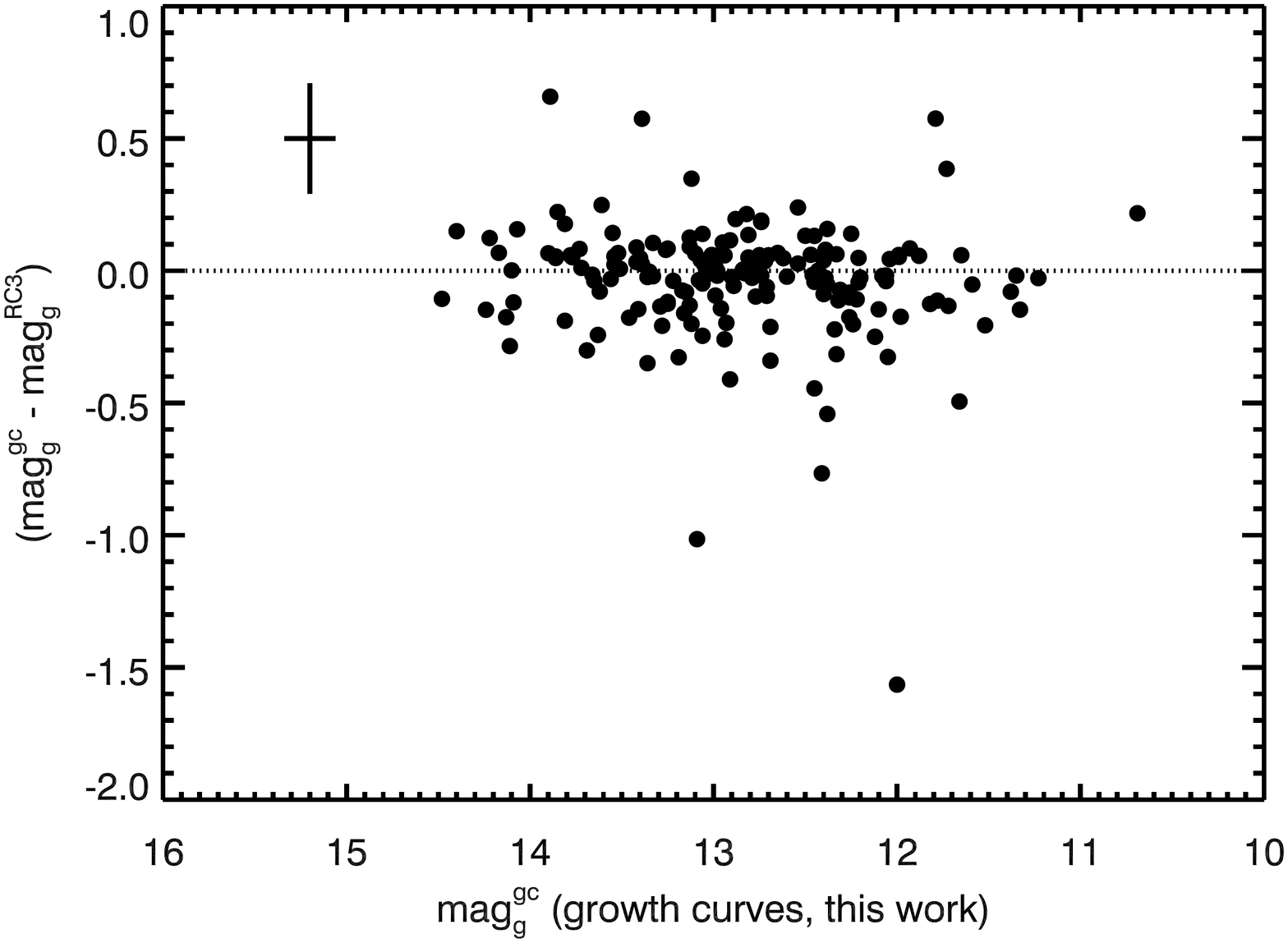}}
\end{center}
\caption[mag_gc_sdss]{Left panel: Comparison of apparent magnitudes obtained from growth curve measurements with those from the 
SDSS pipeline ($\mathit{petroMag}_{\mathrm{r}}$). Right panel:  Comparison of apparent $g$ magnitudes obtained 
from growth curve measurements with estimates of the same derived from the RC3. Both panels show typical error bars in the upper left corner. }
\label{f:mag_gc_sdss}
\end{figure*}

The left panel of 
Figure \ref{f:mag_gc_sdss} shows this comparison between the $r$-band growth curve magnitudes and $\mathit{petroMag}_{\mathrm{r}}$ 
from the SDSS pipeline. There is an overall correlation between the two quantities, which is satisfactory. But clearly, 
growth curve magnitudes are systematically brighter, and more so for bright galaxies. This is naturally explained 
by considering that GC magnitudes are meant to include \emph{all} the flux of the galaxies, whereas Petrosian magnitudes 
have been defined to include a well-defined fraction of the total galaxy flux, as independent as possible of magnitude.  
The correlation of magnitude difference with absolute magnitude is due to the correlation of absolute magnitudes with 
morphological type and therefore S\'ersic $n$ in the CALIFA MS. Indeed, \citet{blanton01} 
show that Petrosian magnitudes contain between 82\% and 100\% of the flux for a de Vaucouleurs and exponential profile, respectively.
The mean difference between the two measurements is $\Delta(\mathit{mag})$ = 0.34 in the sense that 
growth curve magnitudes are brighter.  For correction onto the CALIFA GC system, the offsets that have to be applied per SDSS magnitude 
intervals are: $\mathit{petroMag}_{\mathrm{r}}$ $>14$ : $-0.19$, $14>$ $\mathit{petroMag}_{\mathrm{r}}$ $>13$ : $ -0.22$, 
$13>$ $\mathit{petroMag}_{\mathrm{r}}$ $>12$ : $-0.34$, $\mathit{petroMag}_{\mathrm{r}}$$<12$ : $-0.45$. 
There are a few `catastrophic' outliers, which are due to shredding of large objects in the SDSS 
pipeline. Otherwise the scatter around the mean difference is 0.24 mag. Note that the uncertainty on the magnitudes 
of the CALIFA sample as determined by the SDSS pipeline is of 0.03 mag, which seems very low in light of this comparison. 

There are 172 galaxies in common between the RC3 and the CALIFA MS that have RC3 total magnitudes. 
The right panel of Figure \ref{f:mag_gc_sdss} shows a comparison between the $g$-band growth curve magnitudes and an estimate 
of the $g$ band magnitude determined from the RC3 using their $B$-band total magnitude and $B-V$ colour as well as the following 
equation from \citet{jordi06}: 
\begin{equation}
g-B   =  - 0.370*(B-V) - 0.124
\end{equation}
The mean offset between $g_{\mathrm{GC}}$ and $g_{\mathrm{RC3}}$ is just $-0.04$ mag, with most of the offset due to very few outliers (the median 
difference is $-0.01$ mag). The scatter around the mean difference is 0.22 mag. The mean uncertainty on the RC3 magnitudes is 0.16 magnitudes. 
Together with the 0.14 mag uncertainty on the growth curve measurements, this scatter thus seems mostly due to uncertainties in the determination of the 
total magnitude.

We conclude that reliable photometry of galaxies of the mother sample is now available in the form of these growth curve magnitudes. A systematic 
study of the dependence of flux recovery in the SDSS as a function of galaxy size on the sky and structural properties is, however, beyond 
the scope of the current paper. 

\subsection{Absolute magnitudes and stellar masses}
\label{s:absmag}

To derive absolute magnitudes and stellar masses, one needs to determine the rest-frame SED of the galaxy and convolve it 
with the known filter response functions or multiply with the fitted mass-to-light ratio. Many assumptions and technical tricks 
go into these derivations \citep{walcher11}, and it is beyond the scope of this paper to describe in detail how these are addressed. 
We therefore calculated stellar masses using two existing and well-tested codes, namely \texttt{kcorrect} \citep{blanton07} and 
an algorithm that has been extensively used and tested in \citet[][W08]{walcher08}. Both codes rely on \cite{bruzual03} 
stellar population models with a Chabrier stellar initial mass function \citep{chabrier03}, but the W08 code employs an unpublished 
updated version of the BC03 models, which is termed CB07 (see W08 for details). The codes differ notably in their 
assumptions about the underlying star formation histories, and in their routines to derive the best matching physical properties. 
In particular, the W08 code uses a Bayesian method to derive probability density functions for the 
output parameters, thereby allowing accurate determinations of uncertainties. Both codes sample wide ranges of star formation 
histories (with differences in the details) and dust attenuation amplitudes. We applied both codes only to the 
optical growth curve photometry. Stellar masses agree very well, with a systematic deviation of 0.1 dex in the sense that the W08 
masses are lighter as expected due to the inclusion of secondary bursts in the library of star formation histories (see W08). The 
RMS scatter of 0.15 dex is nearly indistinguishable from the mean 1$\sigma$ uncertainty of 0.11 dex calculated by the W08 code. 
Both codes are equally affected by IMF uncertainties, which may be of the same order as the quoted uncertainties.
Owing to the slight differences between the \texttt{kcorrect} and the W08 masses, there does not seem to be a strong reason to prefer 
one over other, although the W08 masses could be more appropriate in those cases where the galaxies 
did experience recent bursts of star formation. 

We also applied the W08 code to SEDs with added GALEX and 2MASS photometry (see Section \ref{s:ancillary}) which 
provide a better constraint on the dust components. In cases where either in the UV or the NIR photometry 
data points were flagged as bad, these were not used and we reverted to simple optical masses. This makes the final catalogue 
somewhat inhomogeneous. Nevertheless, overall the derived masses are lower by 0.13 dex than the optical ones, 
with a scatter of 0.13 dex. Quoting from W08, `the mean ratios of masses determined without NIR data to the masses derived 
with NIR data are 2.8, 1.50, 1.0 for bins of specific star formation rate log(SSFR/yr$^{-1}$) of $[-16,-13],[-13,-10],[-10,-8]$, respectively.' 
This effect is thus expected. We adopt stellar masses based on the UV, optical and NIR SEDs from now on. 

\begin{figure*}[tbp]
\begin{center}
\resizebox{0.49\hsize}{!}{\includegraphics[]{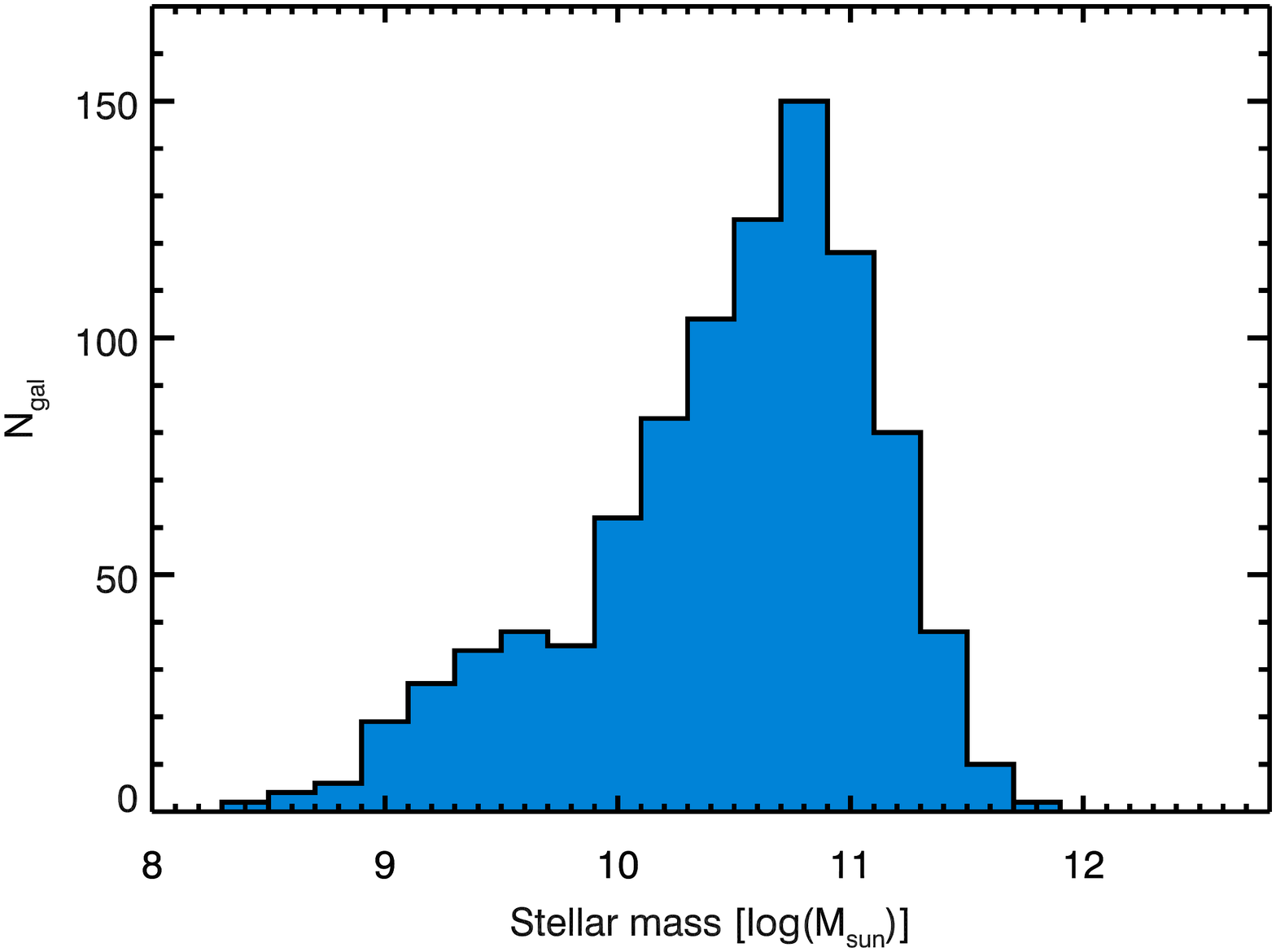}}
\resizebox{0.49\hsize}{!}{\includegraphics[]{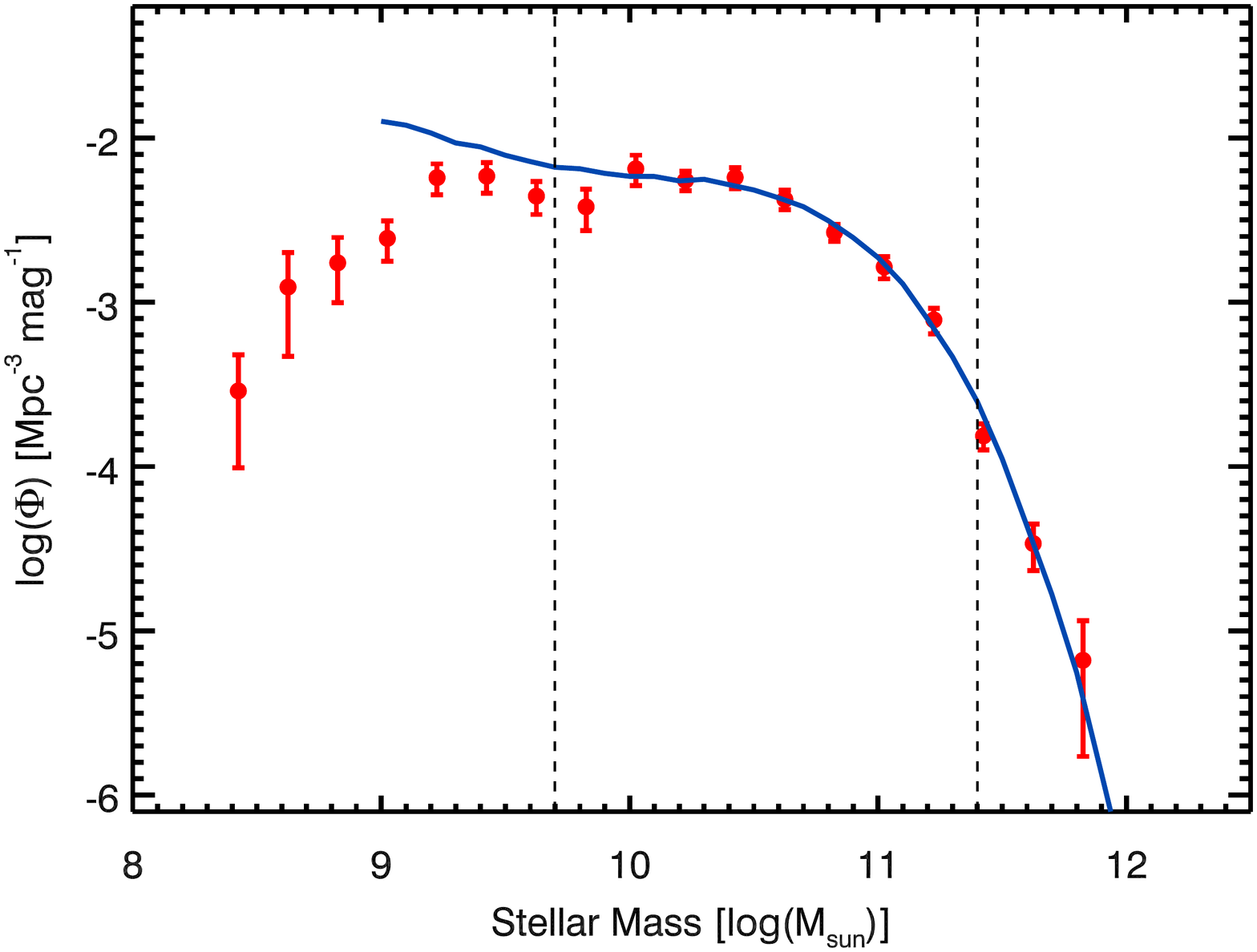}}
\end{center}
\caption[masshist]{\emph{Left panel:}The distribution of stellar masses in the CALIFA sample from a fit to the optical spectral energy distribution. 
\emph{Right panel:} The mass function of the CALIFA mother sample compared with the mass function from \cite{moustakas13}. The two vertical 
lines indicate the representativity limits derived using the same method as in Section \ref{s:selim} from the low-$z$ SDSS comparison sample: 
$9.65 < \log (M^\star/\Msun) < 11.44$.}
\label{f:masshist}
\end{figure*}

Figure \ref{f:masshist} shows the derived stellar mass histogram. The CALIFA sample covers galaxies between 
$10^9$ and $10^{11.5}$ \Msun, with a sharp peak between $10^{10}$ and $2 \times 10^{11}$ \Msun. 
This figure thus shows the range of stellar masses where the statistical power of CALIFA is best. 
Figure \ref{f:masshist} also shows the mass function derived from these stellar masses and the volume corrections 
derived above and compares it with the mass function from \cite{moustakas13}. The near perfect agreement 
over a large range of stellar masses shows the range of stellar masses where the CALIFA 
sample can be used to derive statements about the general galaxy population. 

\subsection{Morphological composition of the sample}
\label{s:morph}

One of the defining characteristics of the MS is that it contains galaxies of all morphological types. When looking through 
the morphological classifications available from public databases we found that these were incomplete for 
our sample \citep[e.g.~Galaxy Zoo 2, 535 matches][]{willett13} or missing a consistent classification in Hubble subtypes (NED). 
We therefore undertook our own reclassification. 

To obtain a morphological classification for the CALIFA galaxies we used human by-eye classification. Five 
co-authors classified all 939 galaxies in the MS according to the following criteria: 
\begin{enumerate}
\item E or S or I for elliptical, spiral, irregular
\item 0-7 (for Es) or 0, 0a, a, ab, b, bc, c, cd, d, m (for S) or r (for I)
\item B for barred, otherwise A. AB if unsure.
\item Merger features, yes or no
\end{enumerate}
For mergers, columns 1 to 3 were filled with the properties of the main object, if possible. If nothing at all was possible U (unknown) was written there. 
The classifiers gave equal weight to SDSS postage stamps in $r$ and $i$ band.

\begin{figure}[tbp]
\begin{center}
\resizebox{1.0\hsize}{!}{\includegraphics[]{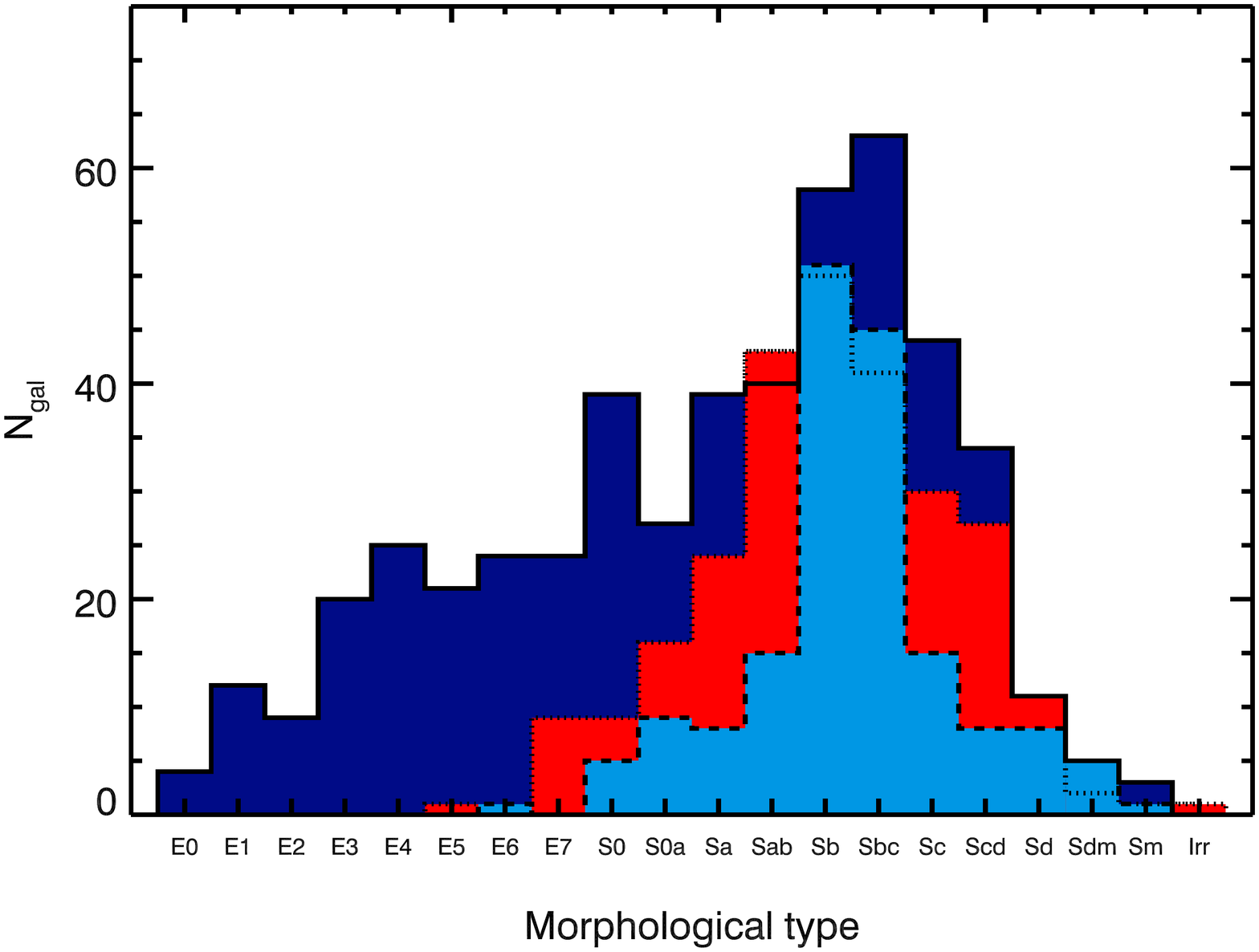}}
\end{center}
\caption[morphhist]{The distribution of morphological types in the CALIFA sample from our own classification. Independent histograms 
are drawn by bar classification as non-barred (meanbar = A, full line, dark blue), strong bar (meanbar=B, dashed, blue) and weak bar 
(meanbar = AB, dotted, red).  }
\label{f:morphhist}
\end{figure}

The five tables obtained were combined, clipping outlier measurements in the calculation of the mean, but keeping them as 
minimum and/or maximum values. Figure \ref{f:morphhist} shows the resulting morphology histogram. We verify that 
the CALIFA MS covers a broad range in galaxy morphologies. 

It may be of interest to note that 8 galaxies in the MS are classified as cD galaxies according to NED. 
These are (with cluster name when known): 
NGC0731, NGC1361, NGC2832 (Abell 779), NGC4556 (Tago 41262), NGC4841A (Abell 1656, Coma), NGC4874 (Abell 1656, Coma), 
NGC5444 (Math 1280, 2MASS 845), NGC6021 (Tago 71733).

\section{Environment}
\label{s:env}

Environmental effects are expected to play a significant role in galaxy evolution. 
However, the many physical processes, their varying amplitudes and timescales 
make it observationally difficult to directly quantify the consequences. One of the difficulties is the challenge 
of defining a general measure of environment. In practice, different measures of 
environment will be relevant for different physical effects. With this in mind we decided to provide a 
range of estimations of environment in the present paper. Generally speaking, environmental measures 
differ by the size of the probed volume and by whether they concern themselves with structures in the galaxy
distribution (e.g. isolated, pairs, groups, clusters, etc) or whether they look at the mean density of galaxies 
on a given spatial scale. For a discussion of standard literature methods see e.g.~\citet{gavazzi10}.
The primary aim of the present section is to verify whether we are lacking any particular kind of environment. 
Given our restricted sample size, the very general aim of this paper, and the difficulties of constructing 
appropriate comparison values, it would be pointless here to dissect the sample into subclasses for every environmental 
measure. This will be undertaken in dedicated papers and in relation to specific scientific goals. 

\subsection{Membership to well known structures}

Galaxies aggregate into structures of very different sizes and scales: from isolation to massive clusters. 
Each scale has a different effect on the evolution of galaxies and no clear boundaries can be defined. For this 
reason we determined the membership of CALIFA galaxies to well known catalogues of galaxy aggregates, as 
one way to characterize their environment. 

In a first step we determined the membership of CALIFA galaxies to catalogues of aggregates of a few galaxies, so that all 
the galaxies in a group are clearly identified: the AMIGA catalogue of isolated galaxies \citep[][]{verdesmontenegro05}, 
isolated pairs and triplets of galaxies \citep{karachentsev72, karachentseva87}, and compact groups of galaxies 
\citep{hickson82}. We also include in this list the Virgo Cluster Catalogue \citep[VCC,][]{binggeli85} with background 
source classification from the GOLDMine database \citep[][]{gavazzi03}. Specifically, 3 of the 35 CALIFA galaxies in the Virgo Cluster 
Catalogue are classified as Virgo background galaxies in GOLDMine. Table \ref{crosco} shows the result from a 
cross-correlation of the CALIFA sample with the catalogues listed above.

\begin{table*}
\begin{center}
\caption{Cross-correlation of CALIFA sample with literature catalogues}
\begin{tabular}{lccccc}
\hline
 & AMIGA & Isolated pairs & Isolated triplets & Hickson & Virgo Cluster \\
\hline
Number of Galaxies & 45                  & 69             & 14                & 17                     & 35                 \\
\hline
\end{tabular}
\label{crosco}
\end{center}
\end{table*}

As a second step we cross-correlated the CALIFA sample with the positions of compilations of loose groups and clusters found 
in the literature \citep{white97, aguerri07, hernandezfernandez12, mahdavi01, miller05, popesso07, shen08, garcia93, 
mahtessian10, tago10, crook07, mahdavi04, berlind06}. In this case, as these aggregates of galaxies are defined by density 
peaks of galaxies in the spatial or radial velocity coordinates, the number of galaxies belonging to each aggregate is
uncertain and in most cases the membership to a given aggregate is derived only for the most massive galaxies. A more 
natural way to ascertain the membership of a galaxy to an aggregate is a combination of its projected distance to the center 
and its relative radial velocity with respect to the systemic radial velocity of the aggregate.

Thus for each CALIFA galaxy we computed the projected distance to the center of all the groups/clusters in units of the virial 
radius. We adopted $R_{200}$, computed following \cite{finn05}, as a good estimate for the virial radius. We also obtained 
the differences in radial velocity with respect to those of the groups/clusters in units of the velocity dispersion. Table~\ref{t:gru_clu} 
contains the results of the cross-correlation of the CALIFA sample with the groups/clusters catalogues previously mentioned. 
The parameters $v_a, \sigma_a$ describe the association, while $v_g$ refers to each galaxy. $D_p$ is the projected distance on sky.

As a way to distinguish between the environments of CALIFA galaxies, we decided 
to separate them into galaxy aggregates with $\sigma \leq 550$~km~s$^{-1}$ (hereafter LV associations), the 
less massive, and those with $\sigma > 550$~km~s$^{-1}$ (hereafter HV associations), the more massive and 
dense \citep{poggianti06}. Note that this separation is purely arbitrary and does not necessarily imply 
a scale of physical transformation. Indeed transformation of satellites may occur at lower $\sigma$ and 
at $M_{\mathrm{halo}} < 10^{13}$ \Msun \citep[e.g.][]{de-lucia12}. 
Given that we are dealing with a very large number of LV and HV associations and that a 
detailed dynamical analysis of all of them is out of the scope of this work, we present the number of CALIFA 
galaxies belonging to an LV/HV association following three different criteria which are usually found in the literature:

\begin{itemize}
\item The number of galaxies that a projected distance lower than the virial radius and with 
$|v_{\mathrm{rad}} - v_{\mathrm{assoc}}| < \sigma_{\mathrm{assoc}}$ of a given LV and HV galaxy association. This criterion identifies the 
members of the cluster core and does not take into account the members from the infall regions or rebounding 
members after a high velocity passage close to the cluster center.
\item The number of galaxies at a projected distance lower than the virial radius and with 
$|v_{\mathrm{rad}} - v_{\mathrm{assoc}}| < 3 \times \sigma_{\mathrm{assoc}}$ of a given LV and HV galaxy association. This criterion includes 
some information about new infalling members but may include some foreground/background members especially close to the 
$R_{200}$ border because of the relaxed $3 \times \sigma$ condition.
\item The number of galaxies falling inside the average caustics proposed by \citet{rines03} for their sample 
of nearby clusters up to a projected distance of $5 \times R_{200}$ from any LV or HV association. This criterion 
seems to be the most appropriate to determine the membership of a galaxy to a LV/HV association because it 
takes into account galaxies from infall regions and also rebounding galaxies. However deviations of the average 
caustics used in this work with respect to the true caustics could lead to incorrect assignments of the individual 
galaxies to LV or HV associations.
\end{itemize}

\begin{table*}
\caption{Summary of the membership of the CALIFA sample to galaxy associations}
\begin{tabular}{ccc|ccc}
\hline
\multicolumn{3}{c}{Belong to a LV association} & \multicolumn{3}{c}{Belong to a HV association} \\
$D_p < R_{200}$ & $D_p < R_{200}$ & $D_p < 5 \times R_{200}$ & $D_p < R_{200}$ & $D_p < R_{200}$ & $D_p < 5 \times R_{200}$ \\
$| v_g - v_a| < \sigma_a$ & $| v_g - v_a| < 3 \times \sigma_a$ & within caustics & $| v_g - v_a| < \sigma_a$ & $| v_g - v_a| < 3 \times \sigma_a$ & within caustics \\
\hline
194 & 387 & 567 & 33 & 70 & 126 \\
\hline
\end{tabular}
\label{t:gru_clu}
\end{table*}

In summary, 246 galaxies likely belong to no known association, 567 likely belong to a low mass association and 126 
likely belong to a high mass association. We conclude that we sample all types of group memberships within the 
CALIFA MS. The sky region covered by the survey includes well known structures such as the Coma/A1367 
supercluster as well as isolated galaxies in the Great Wall. Concerning the Virgo cluster, the lower redshift cut at 1500 km/s implies that 
it is only partly covered by our survey. Virgo has a 3D structure with the main body at $\approx 1200$ km/s, but some subclusters 
further away \citep[2000 km/s, see e.g.~][]{gavazzi99}. Figure \ref{f:env1} (right panel) shows the distribution of galaxies over stellar mass 
and velocity dispersion of their host structure, as well as their morphological type. 

\subsection{Halo mass catalogue}

We matched the CALIFA MS
with the group catalogue extracted from the SDSS DR7 by \citet{wang11} and \citet{yang07}. 
This catalogue uses a group finder and SDSS DR7 to determine group membership and likely 
halo masses for SDSS galaxies. The matching was done by imposing that the angular distance between a CALIFA 
galaxy and a catalogue object be smaller than 1.5\arcsec. This results in 513 CALIFA matched galaxies. The maximum 
angular distance is 1.1\arcsec, and about 50\% of the galaxies have angular distances $\sim$0.1\arcsec.
Besides the mass of the parent halo, the matching also produces information on 
the group hierarchy: A rank of 1 indicates that the galaxy is a central galaxy (the most
massive one of the group), while a rank = 2 labels the galaxy a satellite.

The CALIFA MS contains galaxies belonging to halo masses between $10^{11}$ and $10^{14}$ \Msun. 

\subsection{Local density of the CALIFA galaxies}

The number density of local  galaxies was computed using the projected
comoving  distance to  the $N$th  nearest neighbour  ($d_{N}$)  of the
target galaxy. Thus, the projected galaxy density is defined as 

\begin{equation}
\Sigma_{N}=\frac{N}{\pi (d_{N})^2}.
\end{equation}

We defined the nearest  neighbours using two different samples. First,
we select only those galaxies with spectroscopic redshift located in a
velocity range of $\pm$1000 km  s$^{-1}$ from the target galaxy and with
a  luminosity contrast  of $\pm$2 mag.  These two constraints are
similar to those used by \cite{balogh04a, balogh04b} and allow us
to limit  the contamination by background/foreground  galaxies even if
we  are working  with  projected distances. Secondly,  we defined  a
photometric sample and we  select only those galaxies with photometric
redshifts in  the interval $\mathit{pz} < z_{\mathrm{gal}}+0.1$  to
account for the uncertainties  in the photometric redshifts \citep[e.g., see for a
similar approach][]{baldry06}. To account for possible edge effects in our sample,
we flagged those galaxies with $d_{\mathrm{N}}$ greater than the distance to the
edge of the survey, as these galaxies will have much more uncertain 
environmental densities.

We calculated  the number density  using the third, fifth, eighth, and tenth nearest
neighbours, for both the spectroscopic  and photometric samples. 
The last two measurements were averaged and the differences give
us an indication  of the  uncertainties in the calculated densities
\citep{baldry06}. The mean uncertainty in the same parameter over the 
sample is $\approx 1.4$ galaxies/Mpc$^2$. Another  way  of  testing  the  accuracy  
of  our densities is by  comparing our different estimations based  on the 
number of  neighbours. We found a  good agreement,  
with typical standard deviations of 0.3, 0.4, and  0.5 dex in the
$\Sigma_3-\Sigma_5$,   $\Sigma_3-\Sigma_8$,  and  $\Sigma_3-\Sigma_{10}$
differences,  respectively. Comparison of our values with those of \cite{tempel12} 
also shows good agreement. 
We thus conclude that the presented values are robust and eventual 
differences to other measurement methods will be due to physical differences 
between them. 

For orientation, the density as computed from the 5th nearest neighbour varies 
between 0.1 and 55 galaxies per Mpc$^2$ in our sample. Densities of 
$\Sigma_5 <$ 1 Mpc$^{-2}$ correspond to very low-density environments, 
10 Mpc$^{-2}$ > $\Sigma_5 >$ 1 Mpc$^{-2}$ correspond to loose groups, and 
$\Sigma_5 >$ 10 Mpc$^{-2}$ correspond to compact galaxy groups and clusters \citep[compare][]{aguerri09}. 
According to these criteria, 240 galaxies are located in low density environments, 387 in medium density 
and 310 in high density environments. We thus conclude that the CALIFA MS samples also all environmental densities. 

\subsection{Tidal forces}

To characterize the influence of close neighbours on the CALIFA sample galaxies we followed the 
method by \cite{varela04}. The Varela tidal perturbation $f$ parameter measures the ratio between the internal 
forces ($F_{\mathrm{int}}$) and external tidal forces ($F_{\mathrm{ext}}$) at the outskirts of a given galaxy, as 
caused by satellite/neighbouring galaxies. It does not take into account the relative velocities, as this information 
is not available for most of the non-CALIFA galaxies in the neighbourhood. Relative velocities may have an 
influence on the strength of observed features through the duration of the tidal encounter.

To determine $f$ we searched for local neighbours of each CALIFA sample galaxy in the 
SDSS DR8. For robustness we only extracted galaxies with well determined magnitudes and Petrosian 
radii in the $r$ and $g$ band. For each galaxy the information is taken from the catalogues PhotoObj and SpecObj. 
The criteria to find the satellites were:

\begin{enumerate}
\item Objects classified as galaxies (type=3). 
\item Up to 200 kpc from the CALIFA target (assuming simple redshift-based radial and tangential distances)
\item With reported values of Petrosian radii at 90\%  and at 50\% flux in $r$ and $g$ bands
(acceptable values, we excluded objects with negative errors in the Petrosian radii).
\item With sizes of at least 2 kpc (as provided by the $\mathit{petroRad}_{\mathrm{r}}$).
\item With good quality flags [flags$_r$=0 AND flags$_g$ = 0]
\end{enumerate}

We then calculated for all identified neighbours of a CALIFA galaxy 
the $f$ parameter according to \cite{varela04}, i.e. the tidal force exerted by the 
neighbour onto the CALIFA galaxy. Using $m_{\mathrm{G}}$ and $m_{\mathrm{P}}$ as the apparent magnitudes 
of the primary and perturber galaxies, respectively, $R$ as the size of the galaxy and $D_{\mathrm{p}}$ as the 
projected distance between the galaxy and the perturber on the plane of the sky at the distance of the primary, 
the equation given by \cite{varela04} to calculate $f$ is: 

\begin{equation}
f = \mbox{log}\left( \frac{F_{\mathrm{ext}}}{F_{\mathrm{int}}}\right) = 3 ~ \mbox{log}\left( \frac{R}{D_{\mathrm{p}}}\right) + 0.4~  (m_{\mathrm{G}} - m_{\mathrm{P}})
\end{equation}

Conceptually, an $f$ value below $-4.5$ indicates no tidal influence, values between $-4.5$ and $-2$ indicate 
that there is weak influence at most and objects with an $f$ value above $-2$ could be producing 
interaction effects on the CALIFA galaxy. 
From this environmental measure alone we would conclude that 335 galaxies 
in the CALIFA MS are completely isolated. On the other hand, there are 185 that 
could suffer from strong tidal effects. 
These numbers again confirm that the CALIFA sample is suitable for studying 
effects of galaxy interactions, while simultaneously providing a bona fide comparison sample of 
completely isolated galaxies. 

\subsection{Interactions}

In this catalogue, our goal was to select from the MS those galaxies with evident signatures of interaction/merging 
(i.e., tails, bridges, rings, etc. ). Three classifications were performed in total by different members of the 
collaboration with different scientific goals. 

\begin{enumerate}
\item The interaction flag described in Section \ref{s:morph}. 
\item SDSS images were inspected for features indicating interaction. The unsharp masking 
technique was used. We use here a binary flag, lumping together all different morphological hints 
of interaction (streams, disc-disturbance, compact group membership) into an interaction flag, all others being `non-interacting'. 
\item Yet another independent by-eye classification on the SDSS images was performed, again classifying galaxies in a binary flag 
as interacting or non-interacting. While the technique is the same as in point 1., the classifiers were entirely independent. 
\end{enumerate}

The final number of interacting galaxies was determined by collecting those galaxies that were flagged as interacting in 
two of the three previous catalogues. The total number of visually interacting galaxies in the MS is 152, 
approximately the same as in the previous section.  

\subsection{Results on Environment}

\begin{figure*}[tbp]
\begin{center}
\resizebox{0.49\hsize}{!}{\includegraphics[]{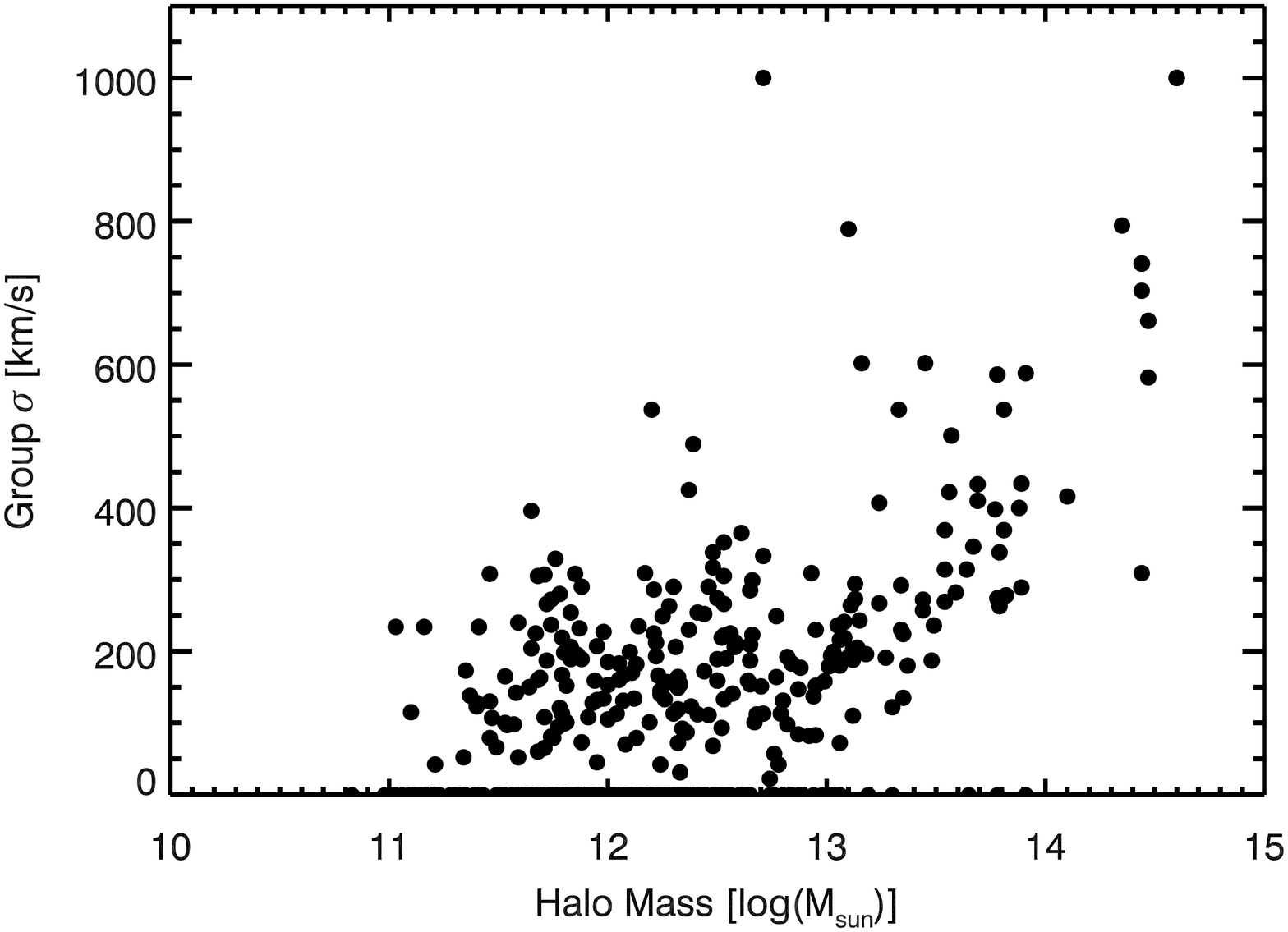}}
\resizebox{0.49\hsize}{!}{\includegraphics[]{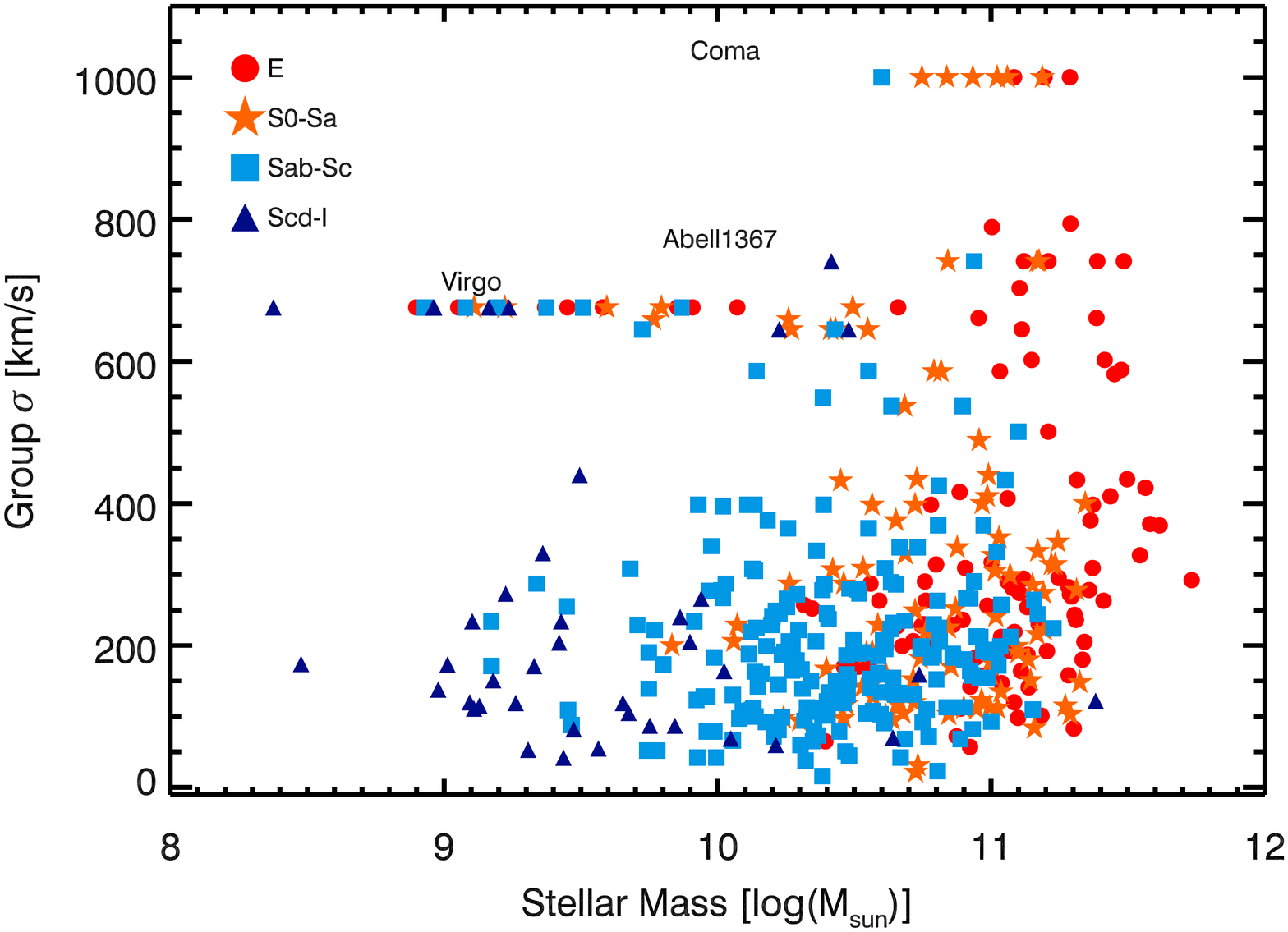}}
\end{center}
\caption[env1]{\emph{Left panel:} Comparison between a pair of related environment measures. The mass of the host halo 
from the \cite{wang11} catalogue vs. the velocity dispersion of the closest known structure compiled in this work. \emph{Right panel: }
Stellar mass distribution of members in known structures. The largest known structures have been identified. Galaxies have been 
colour coded by their morphological classification. }
\label{f:env1}
\end{figure*}

All presented environment measures are useful for different physical questions, and not all of them are 
actually related to each other. Figure \ref{f:env1} shows as an example that while two of the global environment 
measures roughly agree, there seem to be differences in the details which may either represent measurement 
uncertainties or physical differences between the two measures. It is beyond the scope of the present paper 
to solve this question; we will be address the issue in future CALIFA papers. 

Two local parameters that should be closely related are the Varela $f$ parameter and the interaction state of a galaxy. 
We find that for non-interacting galaxies the $f$ parameter is $-4.0$ with a standard deviation of $1.7$, while for interacting 
galaxies it is $-2.9 \pm 2.0$. Clearly, while the difference in the mean indicates some correlation between the Varela $f$ 
parameter and the interaction state, the distributions of interacting and non-interacting galaxies in $f$ overlap 
significantly. Galaxies can be in different stages of interaction, which may or may not be associated to 
visible signs of interaction. Thus again, these two physically different measures of local environment only show a 
weak correspondence. It will be interesting to use the CALIFA velocity fields to probe the influence of interactions 
in more detail, in particular for outlier galaxies, i.e. those with a large $f$ parameter, but no sign of interaction from 
optical imaging and those with optical signs of disturbances but a small $f$ parameter. 

CALIFA galaxies represent all ranges of environment, high and low galaxy densities, high and low halo/group 
masses, and isolation vs. interaction. The CALIFA sample is thus well placed to provide interesting insights on 
the environmental effects in galaxy evolution.

\section{AGN Content}

It would be certainly of interest if CALIFA could be used to scrutinize `active' and `inactive' galaxies at the same time. Unravelling the  AGN 
content in the CALIFA sample from the full integral field spectroscopy will be the subject of a separate paper. For the 
present work we limit ourselves to a quick assessment of the evidence for AGN using the data prior to performing any 
CALIFA observations. AGN can potentially be identified via several independent methods, some of which are used in the 
following. 

\subsection{Classical emission-line diagnostics}

\begin{figure}[t]
\resizebox{\hsize}{!}{\includegraphics{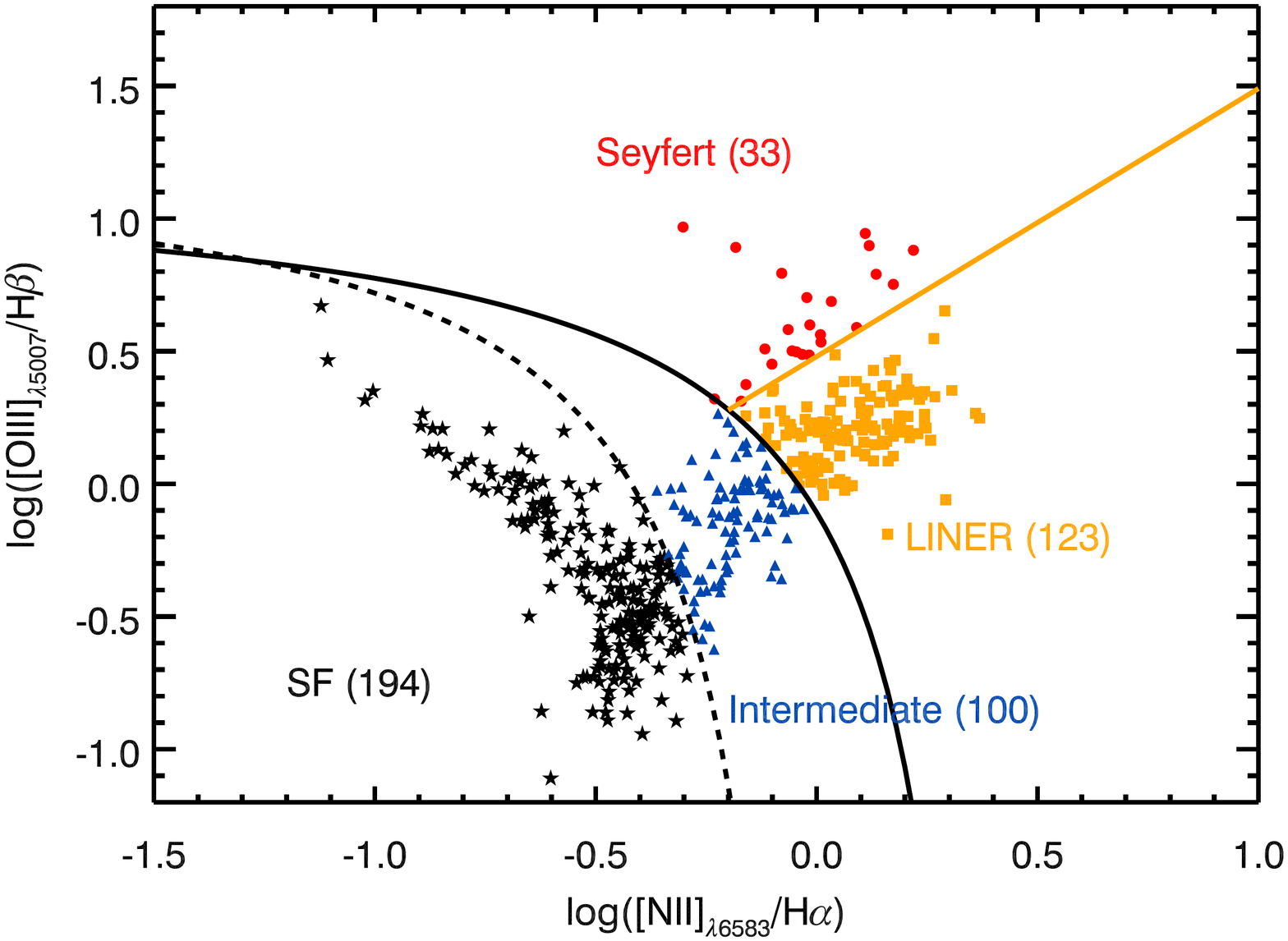}}
\caption{Standard emission-line diagnostic diagram for CALIFA galaxies with SDSS spectra. Only spectra centered on the galaxy nucleus 
($<$3\arcsec) and with S/N$>$3 in all emission lines are shown here. The demarcation lines by \citet{kewley01} (black), 
\citet{kauffmann03b} (black dotted), and \citet{cid-fernandes10} (yellow) are used to characterize the objects into star forming 
(black), Seyferts (red), SF/AGN intermediates (blue), and LINER-like (orange) objects. The number of objects per class is indicated on the plot. }
\label{fig:SDSS_BPT}
\end{figure}

The emission-line fluxes for all SDSS spectra of DR7 were measured and provided by the MPA-JHU 
group\footnote{www.mpa-garching.mpg.de/SDSS/DR7/} as value-added catalogues following the method outlined 
in \cite{tremonti04}. Here we use the classical [\ion{O}{III}]\,$\lambda5007$/H$\beta$ vs. 
[\ion{N}{II}]\,$\lambda6583$/H$\alpha$ diagram introduced by \citet{baldwin81} to discriminate between different 
ionization sources at the galaxy centre of CALIFA galaxies. We use the demarcation lines of \citet{kauffmann03b}, 
\citet{kewley01} and \citet{cid-fernandes10} to classify the objects into star forming (SF), Seyferts, SF/AGN intermediate, 
and LINER-like galaxies. Of 582 galaxies which have an SDSS spectrum centered 
within 3\arcsec of the nucleus, 450 have S/N$>3$ 
in all used emission lines and those are shown in Figure \ref{fig:SDSS_BPT}. The other 132 (22\%) have at least one emission 
line that is too faint for a reliable classification. We find that 194 of 450 galaxies (43\%) are clearly dominated 
by star formation, 100 objects  (22\%) are in the intermediate zone between the SF and AGN branches, 24 
objects (5.3\%) are of Seyfert type and 132 galaxies (29\%) have LINER-like emission-line ratios. 

\subsection{X-ray luminosities}

When the X-ray luminosity of a source exceeds $\log(L_{2-10\,\mathrm{keV}}/[\mathrm{erg\,s}^{-1}])>42$ \citep[e.g.][]{szokoly04} in 
the soft or $\log(L_{15-195\,\mathrm{keV}}/[\mathrm{erg\,s}^{-1}])>42.2$ in the hard band, it is most likely harbouring an AGN. A 
large fraction of soft X-ray photons are absorbed in obscured (type 2) AGN, so the ROSAT all-sky survey \citep{voges99} does 
not efficiently identify the low-luminosity type 2 AGN that dominate the AGN population in CALIFA. Instead, we matched the 
CALIFA galaxies with the Swift BAT 70-month hard X-ray survey \citep{baumgartner12} that contains 15 confirmed counterparts with 
$\log(L_{15-195\,\mathrm{keV}}/[\mathrm{erg\,s}^{-1}])>42.2$ clearly indicative of AGN. This hard X-ray sample includes Mrk 79, 
which is a well-known type 1 AGN that is part of the CALIFA sample. 

\subsection{The incidence of radio AGN}

\begin{figure}[h]
\resizebox{\hsize}{!}{\includegraphics{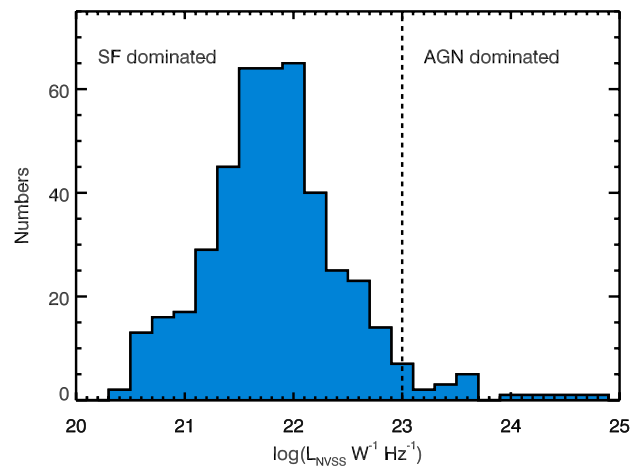}}
\caption{Histogram of NVSS continuum radio luminosities at 1.4\,GHz for all detected CALIFA objects.}
\label{fig:NVSS_hist}
\end{figure}

A completely different signature of nuclear activity is the jets released from the AGN that can be identified by their enhanced 
radio emission. Ongoing star formation usually produces also some level of radio emission, but the number density of 
radio-AGN dominates over that of star forming galaxies above a radio continuum luminosity of 
$L_{1.4\mathrm{GHz}}>10^{23}\mathrm{W}\,\mathrm{Hz}^{-1}$ \citep{best12}. The radio luminosity 
distribution of CALIFA galaxies detected by the NVSS survey is shown in Figure \ref{fig:NVSS_hist}. We identified 15 galaxies with  
$L_{1.4\mathrm{GHz}}>10^{23}\mathrm{W}\,\mathrm{Hz}^{-1}$ in the CALIFA MS for which we carefully checked the radio morphology 
in the corresponding NVSS and FIRST images, when available. Removing the known ULIRG Arp220, 11 of these show either clear jet-like 
structure or are hosted in bulge-dominated galaxies for which strong star formation is not expected.



For galaxies with $L_{1.4\mathrm{GHz}}<10^{23}\mathrm{W}\,\mathrm{Hz}^{-1}$ it is difficult to identify the origin of 
the radio emission without additional indications. Here we used the criterion introduced by \citet{best05}, incorporating 
the 4000\AA\ break strength ($\mathrm{D4000}_N$ index). The 4000\AA\ break strength was taken from the analysis of 
the SDSS DR7 spectra which reduced the sample again to 582 objects. Because the 3\arcsec\ apertures of the SDSS 
fibres cover only the central part of the galaxy we excluded all radio sources with extended emission 
that follows that of the host galaxy. 
With the 4000\AA\ break strength as an age indicator of the stellar population we can identify 17 additional potential radio AGN. These are 
predominantly located in elliptical or lenticular galaxies that display no significant emission lines or LINER-type spectra. 
The only object in common between the three type 1 AGN samples here is NGC4874, which is a cD galaxy detected in the radio and the x-rays. 

This short accounting shows that based on the evidence available prior to the CALIFA spectroscopy, the AGN fraction 
in CALIFA galaxies appears to be around 6\%. Even if this low fraction should be confirmed there will still be approximately 
30 AGN host galaxies for which the CALIFA data will provide detailed insights.

\section{Summary}

This paper is devoted to a detailed description of the CALIFA mother sample (MS). The main feature of the CALIFA sample is that it has been selected 
by diameter to fill the field of view of the IFU. The observed sample will be a randomly selected subset 
of the MS and will thus share all its properties, albeit with somewhat reduced statistical power. To fully 
characterize the sample, we have derived or collected a number of properties for its galaxies, in particular integrated optical 
magnitudes, stellar masses, and five different environmental measures, and we have identified AGN. 
Secondary data products derived in the context of this paper will be made available on the webpage of the 
CALIFA survey (http://www.caha.es/CALIFA/). 

We conclude the following:
\begin{itemize}
\item The MS is representative for the general galaxy population with the following limits: $-19.0$ to $-23.1$ in $r$-band 
absolute magnitude, 1.7 to 11.5 kpc in half light radii, and 9.7 to 11.4 in log(stellar mass/\Msun). 
\item Below $M_{r,\mathrm{p}} = -19$, the MS contains mostly edge-on galaxies elevated into the sample by projection 
effects acting on their half-light major axes. Above $M_{r,\mathrm{p}} = -23.1$ the CALIFA sample is limited by the total available volume in the 
sense that such luminous galaxies are very rare and thus are not represented in the volume available within our redshift limits. 
\item The application of volume corrections allows the derivation of space densities and distribution functions of any measurable galaxy 
physical property from the CALIFA sample. We have derived the necessary corrections for the local underdensity of the Universe. 
\item More than $97$\% of CALIFA galaxies are covered out to more than $2 \times r_{\mathrm{50}}$ at a typical spatial resolution of 1 kpc in the mean. 
\item The sample covers all environments, from field galaxies to cluster environments, from isolated to interacting to merging galaxies
\item The sample contains few easily identified, luminous AGN as these are rare in the local galaxy population. Nevertheless, the final observed 
sample will contain approximately 30 Seyfert galaxies. 
\end{itemize}

Future IFS surveys should feel encouraged to consider diameter selection, as it provides an efficient use of the field of view and 
-- as we have shown in this paper -- leads to a controlled sample with benign properties. While future IFS surveys will probably be 
superior to CALIFA in sample size, we stress that CALIFA will occupy a unique place in parameter space for a long 
time to come in its combination of field of view, spatial resolution and S/N, which larger surveys will struggle to match due to limits imposed 
by spectrograph and detector sizes, i.e. by the number of resolution elements. 

\begin{acknowledgements}

This study makes uses of the data provided by the Calar Alto Legacy Integral Field Area (CALIFA) survey (http://califa.caha.es). CALIFA is the first legacy survey being performed at Calar Alto. The CALIFA collaboration would like to thank the IAA-CSIC and MPIA-MPG as major partners of the observatory, and CAHA itself, for the unique access to telescope time and support in manpower and infrastructures.  The CALIFA collaboration thanks also the CAHA staff for the dedication to this project. \\

We thank Mike Blanton for helpful discussions on the SDSS survey footprint and for providing the NYU low-$z$ catalogue to the community. CJW acknowledges useful discussion with Nick Scott, Davor Krajnovic and Remco van den Bosch as well as support through the Marie Curie Career Integration Grant 303912. We thank the anonymous referee for a careful reading of the paper and several suggestions that improved its presentation.  IM acknowledges the financial support from the Spanish grant AYA2010-15169 and from the Junta de Andalucia through TIC-114 and the Excellence Project P08-TIC-03531. RGD, EP and RGB acknowledge support from the Spanish Ministerio de Economia y Competitividad, through projects AYA2010-15081. RAM is funded by the Spanish programme of International Campus of Excellence Moncloa (CEI). JIP acknowledges financial support from the Spanish MINECO under grant AYA2010-21887-C04-01 and from Junta de Andaluc\'{\i}a Excellence Project PEX2011-FQM7058. Support for LG is provided by the Ministry of Economy, Development, and Tourism's Millennium Science Initiative through grant IC12009, awarded to The Millennium Institute of Astrophysics, MAS. LG acknowledges support by CONICYT through FONDECYT grant 3140566. AM-I acknowledges support from Agence Nationale de la Recherche through the STILISM project (ANR-12-BS05-0016-02). AM-I and SB acknowledge support from BMBF through the Erasmus-F project (grant number 05 A12BA1). JMA acknowledges support from the European Research Council Starting Grant (SEDmorph; P.I. V. Wild). AG acknowledges funding from the European Union Seventh Framework Programme (FP7/2007-2013) under grant agreement n. 267251. KS acknowledges financial support from the Natural Sciences and Engineering Research Council of Canada (NSERC). BJ acknowledges support by the projects RVO67985815 and M100031201 of the Academy of Sciences of the Czech Republic. 

This research has made use of the NASA/IPAC Extragalactic Database (NED), which is operated by the Jet Propulsion Laboratory, California Institute of Technology, under contract with the National Aeronautics and Space Administration. This research has made use of the SIMBAD database, operated at CDS, Strasbourg, France. We acknowledge the usage of the HyperLeda database (http://leda.univ-lyon1.fr). \\

Funding for the Sloan Digital Sky Survey (SDSS) has been provided by the Alfred P. Sloan Foundation, the Participating Institutions, the National Aeronautics and Space Administration, the National Science Foundation, the U.S. Department of Energy, the Japanese Monbukagakusho, and the Max Planck Society. The SDSS Web site is http://www.sdss.org/. The SDSS is managed by the Astrophysical Research Consortium (ARC) for the Participating Institutions. The Participating Institutions are The University of Chicago, Fermilab, the Institute for Advanced Study, the Japan Participation Group, The Johns Hopkins University, Los Alamos National Laboratory, the Max-Planck-Institute for Astronomy (MPIA), the Max-Planck-Institute for Astrophysics (MPA), New Mexico State University, University of Pittsburgh, Princeton University, the United States Naval Observatory, and the University of Washington.

\end{acknowledgements}

\bibliographystyle{aa}
\bibliography{CALIFAsample}

\end{document}